\documentclass[preprint,12pt]{elsarticle}
\usepackage{graphicx}
\usepackage{dcolumn}
\usepackage{bm}
\usepackage[hidelinks]{hyperref}
\usepackage{academicons}
\usepackage{xcolor}
\usepackage{xcolor}
\hypersetup{
    colorlinks,
    linkcolor={blue},
    citecolor={red},
    urlcolor={blue!80!black}
}
\usepackage[framemethod=TikZ]{mdframed}
\usepackage[most]{tcolorbox}
\usepackage{longtable}
\usepackage{mathcomp}
\usepackage{pifont}
\usepackage{xcolor}
\usepackage{amsthm}
\usepackage{PGF}
\usepackage{pgfpages}
\usepackage{amsmath}
\usepackage{amsfonts}
\usepackage{amssymb}
\usepackage{graphicx}
\usepackage{bm}
\usepackage{hyperref}
\usepackage{lipsum}
\usepackage{array}
\usepackage{booktabs}
\usepackage{tikz}
\usepackage{titlesec}
\usepackage{hyperref}
\hypersetup{bookmarksnumbered}
\def\be{\begin{equation}}
\def\ee{\end{equation}}

\usepackage{amsmath}
\usepackage{empheq}
\usepackage[most]{tcolorbox}
\newtcbox{\mymath}[1][]{%
    nobeforeafter, math upper, tcbox raise base,
    enhanced, colframe=blue!30!black,
    colback=blue!30, boxrule=1pt,
    #1}
\usepackage{mdframed}
\usepackage{lipsum}
\usepackage{PGF}
\usepackage{pgfkeys}
\usepackage{pgffor}
\usepackage{pgfcalendar}
\usepackage{pgfpages}
\usepackage{tikz}
\usetikzlibrary{calc}
\usepackage{array}
\usepackage{booktabs}
\usepackage{multirow}
\usepackage{latexsym}
\usepackage[english]{babel}
\usepackage[most]{tcolorbox}

\newtcolorbox{myquote}[1][]{%
    colback=black!5,
    colframe=black!5,
    notitle,
    sharp corners,
    borderline west={2pt}{0pt}{red!80!black},
    enhanced,
    breakable,
    }
\usepackage{enumitem}

{\begin{itemize}\item[]}
{\end{itemize}}
\usepackage{graphicx}
\usepackage{mathtools}

\usepackage{amssymb}
\begin{document}

\begin{frontmatter}


\title{\textcolor[rgb]{0.00,0.40,0.80}{\bfseries{\boldmath
Quantum Quintom Cosmology
}}}

\author[inst1]{Behzad Tajahmad}
\affiliation[inst1]{behzadtajahmad@yahoo.com}


\begin{abstract}
This work applies the principles of quantum cosmology to examine models incorporating a quintom field. Specifically, three distinct models are analyzed: a simplified toy model, a model featuring an exponential quintom potential, and one where the quintom field is coupled with a negative cosmological constant. For each case, we study the classical trajectories within the configuration space, present solutions to the Wheeler-DeWitt equation in quantum cosmology, and discuss physical interpretations and consequences. A key focus is the behavior of wave packets in the minisuperspace framework. Notably, the correspondence principle (connection between classical and quantum solutions) is also demonstrated. Furthermore, the appropriate quintom duality is introduced, and we discuss its consequences. A section including interesting and challenging discussions is also presented.
\end{abstract}

\end{frontmatter}
\newpage
\tableofcontents
\newpage

\section{Introduction\label{Introduction}}
Current observations of our universe indicate that its expansion is accelerating~\cite{acu}. Despite extensive research, the fundamental nature of the matter responsible for this phenomenon remains unknown, as none of the known forms of matter can account for it. This elusive component, which constitutes a dominant fraction of the universe's energy density, has been termed ``dark energy.'' While the cosmological constant represents the simplest and most widely recognized candidate for dark energy, theoretical physics offers a broader spectrum of possibilities, including scalar fields with specific kinetic and potential energy configurations.

Dark energy is characterized by its negative pressure, which drives the repulsive force behind the universe's accelerated expansion. Within the framework of general relativity, dark energy must violate the strong energy condition, expressed as \( \rho + 3P > 0 \) and \( \rho > 0 \). Adopting a barotropic equation of state \( P = \omega \rho \) (where \( \omega \) is a constant, and \( P \) and \( \rho \) denote pressure and density, respectively), this violation implies \( \omega < -1/3 \). Recent observational data suggest that dark energy increasingly favors more negative values of \( \omega \), with \( \omega \leq -1 \) in the present epoch, whereas \( \omega > -1 \) in the past. This conclusion holds irrespective of whether the universe is assumed to be flat~\cite{q16,q18,q20} or not~\cite{q21}. Such behavior leads to a violation of the null energy condition~\cite{q03}, \( \rho + P > 0 \), as well as other related conditions, including the weak energy condition (\( \rho > 0 \), \( \rho + P > 0 \)) and the dominant energy condition (\( \rho > 0 \), \( -\rho < P < \rho \)). Dark energy exhibiting \( \omega \leq -1 \) is referred to as ``phantom'' energy~\cite{q06,q071,q072}. Phantom fields can be modeled by scalar fields with negative kinetic energy, often called ``ghost fields.'' Although these fields present theoretical challenges~\cite{q08}, they remain observationally viable candidates for dark energy and warrant further investigation. Notably, some phantom field models avoid pathological behavior in the ultraviolet regime~\cite{q010}. Phantom energy leads to a unique cosmological scenario known as the ``big-rip singularity''~\cite{q06,q071,q072,q011}, where energy density and pressure diverge as the scale factor \( a(t) \) grows infinitely within a finite time. This contrasts with the ``big crunch,'' where energy density and pressure become infinite as the scale factor approaches zero. Another intriguing possibility is the ``big brake,'' characterized by a vanishing expansion rate and an infinitely negative acceleration~\cite{q012}. Additionally, other exotic singularities may arise, such as the sudden future singularity~\cite{q013} and its generalized form~\cite{q014}, where higher-order derivatives of the scale factor diverge while the scale factor and energy density remain smooth. Type III~\cite{q015} and type IV~\cite{q016} singularities also exhibit smooth scale factor evolution but differ in their severity compared to the big rip.

A major challenge in modern cosmology is explaining the transition of dark energy across \( \omega = -1 \). In quintessence models with positive potential, \( \omega \) always lies within \(-1 \leq \omega \leq 1\). Conversely, phantom models, characterized by negative kinetic terms, yield \( \omega \leq -1 \). Neither quintessence nor phantom alone can account for a crossing of \( \omega = -1 \). While k-essence~\cite{q27} permits both \( \omega < -1 \) and \( \omega \geq -1 \), ref.~\cite{q39} demonstrates that achieving a smooth transition across \( \omega = -1 \) is highly nontrivial. To address this, the quintom model was introduced in ref.~\cite{q19}, combining an ordinary scalar field with a phantom field. This framework allows \( \omega \) to cross \(-1\) dynamically. Quintom models belong to the broader class of dynamical dark energy scenarios~\cite{q40,q41,q42}, originally proposed to explain cosmic acceleration. However, these models must be rigorously tested across different cosmic epochs to assess their viability as unified cosmological descriptions.
It woth to mention that quintom scenario is in agreement with observational data~\cite{new1,new2}. Interestingly, ref.~\cite{qp004} demonstrates through the minisuperspace approach that the quintom scenario can emerge from \( f(Q) \)-cosmology.
Quintom and other multi-scalar field models have been extensively studied in the literature---see refs.~\cite{q33,q34,q35,q36,q37} for detailed discussions.

Traditionally, the evolution of the universe has been studied through classical cosmology, neglecting quantum effects on cosmic scales. This paper aims to bridge that gap by investigating quantum cosmology in the context of a quintom field. Given the overwhelming experimental and theoretical support for the universality of quantum theory~\cite{q029}, it is natural to expect that the universe as a whole should also be described quantum mechanically. If the quintom field plays a significant role in cosmic evolution, it is crucial to examine whether it introduces deviations from standard quantum cosmology and to explore the resulting physical implications.

Quantum cosmology is fundamentally rooted in a theory of quantum gravity~\cite{q030}. Promising candidates include string theory, loop quantum gravity, and quantum geometrodynamics. Our analysis is based on the Wheeler-DeWitt equation derived from quantum geometrodynamics, a framework expected to remain valid at least below the Planck scale. Near the Planck scale, alternative approaches like loop quantum cosmology~\cite{q031} may become relevant, though such modifications are beyond the scope of this work and left for future study.

A key feature of the Wheeler-DeWitt equation is its local hyperbolic signature~\cite{q030}. In regions of configuration space near closed Friedmann cosmologies, the equation exhibits global hyperbolicity, with the kinetic term containing a single minus sign~\cite{q033}. This negative contribution is associated with the scale factor, effectively allowing it to mimic a phantom field in a certain sense. The indefinite kinetic term arises from the attractive nature of gravity~\cite{q034}. Another defining property of the Wheeler-DeWitt equation is its independence from an external time parameter~\cite{q030}, a consequence of classical reparametrization invariance. Thus, a consistent quantum cosmology must rely on the intrinsic structure of this equation, avoiding ad hoc notions of external time. Classical trajectories should instead be analyzed in a configuration space where the time parameter \( t \) has been eliminated. The hyperbolic nature of the Wheeler-DeWitt equation suggests that boundary conditions should be imposed at fixed values of the scale factor, analogous to wave equations. This is particularly important for constructing wave packets that follow classical trajectories, resembling standing tube-like structures in configuration space~\cite{q035,q037}. Additionally, this framework is essential for interpreting pre- and post-big-bang phases in quantum string cosmology~\cite{q039}.

The inclusion of a phantom field alters the structure of the Wheeler-DeWitt equation. When the phantom field dominates alongside the scale factor, the equation becomes elliptic. In more general cases, it transitions to a mixed (ultrahyperbolic) form, complicating the formulation of boundary conditions. A detailed analysis of quantum phantom cosmology can be found in ref.~\cite{qp}. Furthermore, refs.~\cite{q072,qp002,qp003} explore whether phantom fields can exist in a UV-complete theory.

In this paper, we present the formal framework for quantum quintom cosmology---incorporating both canonical and phantom fields---and examine its key physical implications.
The paper is structured as follows. Section~\ref{sect.2} derives and solves the classical equations of motion for the quintom field in a Friedmann universe. We analyze three specific models: a toy model with vanishing phantom potential, an interacting exponential quintom potential (multiplicative mode), and a cosh-potential (collective mode) with a negative cosmological constant. Section~\ref{sect.3} extends this discussion to the corresponding quantum theory, solving the Wheeler-DeWitt equation exactly for the toy model and the exponential potential case. We construct wave packet solutions and demonstrate that quantum effects become significant near the classical big-rip singularity, effectively resolving it. In realistic scalar field models, the wave function vanishes at the big bang, suggesting its avoidance in quantum theory. In section~\ref{sect.4}, several interesting and challenging discussions are posed. Finally, we summarize our findings and conclusions.
\section{Classical Quintom Cosmology\label{sect.2}}
In this section, we first obtain the classical equations and then consider classical quintom trajectories in three models separately:
\begin{itemize}
  \item No quintom potential (i.e., vanishing quintom potential and vanishing cosmological constant);
  \item Interacting exponential scalar field potential of quintom and vanishing cosmological constant;
  \item Scalar field fluid and negative cosmological constant.
\end{itemize}
The reason for considering classical behavior is that when we acquire quantum solutions in the next section, we have a criterion for taking into account the correspondence principle (leading to classical behavior at $\hbar \to 0$).
\subsection{Classical equations of motion\label{subsect.2.1}}
We consider the Friedmann universe characterized by the scale factor \( a(t) \) and two homogeneous scalar fields: a canonical scalar field \( \varphi_{1}(t) \) and a phantom scalar field \( \varphi_{2}(t) \). In this scenario, we assume that both the canonical and phantom fields dominate significantly over any other matter degrees of freedom, leaving them as the primary dynamical components alongside the scale factor. As a result, the configuration space for the system is described by \( \{a, \varphi_{1}, \varphi_{2}\} \). The action governing this setup is expressed as follows:
\begin{align}\label{action1}
S=\frac{3}{\kappa^{2}}\int \mathrm{d}t\, N\left(-\frac{a \dot{a}^2}{N^2}+\mathcal{K} a-\frac{\Lambda a^{3}}{3} \right)+\frac{1}{2}\int \mathrm{d}t\, N a^{3}\left(
\frac{\dot{\varphi}_{1}^{2}-\dot{\varphi}_{2}^{2}}{N^{2}}
-2V(\varphi_{1},\varphi_{2})
\right),
\end{align}
in which the dot denotes a differentiation with respect to time $t$. In this context, $\kappa^{2} = 8\pi G$, where $G$ represents the gravitational constant, and $N$ denotes the lapse function. The term $\mathcal{K} = 0, \pm 1$ corresponds to the curvature index, indicating whether the spatial geometry is flat ($\mathcal{K} = 0$), closed ($\mathcal{K} = +1$), or open ($\mathcal{K} = -1$). The symbol $\Lambda$ stands for the cosmological constant, while $V(\varphi_{1}, \varphi_{2})$ refers to the potential associated with the quintom fields. For simplicity, the speed of light has been set to $1$.

By assigning $N=1$, the time parameter corresponds to the standard Friedmann cosmic time. Consequently, the action (\ref{action1}) simplifies to:
\begin{align}\label{action2}
S=\frac{3}{\kappa^{2}}\int \mathrm{d}t\, \left(-a \dot{a}^2+\mathcal{K} a-\frac{\Lambda a^{3}}{3} \right)+\frac{1}{2}\int \mathrm{d}t\, a^{3}\left(
\dot{\varphi}_{1}^{2}-\dot{\varphi}_{2}^{2}-2V(\varphi_{1},\varphi_{2})
\right).
\end{align}
The canonical conjugate momenta corresponding to the scale factor and the scalar fields are given by:
\begin{align}\label{momenta}
\Pi_{a}=-\frac{6a\dot{a}}{\kappa^{2}}, \qquad
\Pi_{\varphi_{1}}=a^{3}\dot{\varphi_{1}}, \qquad
\Pi_{\varphi_{2}}=-a^{3}\dot{\varphi_{2}}.
\end{align}
According to (\ref{action2}), the associated Hamiltonian $\mathcal{H}$ reads
\begin{align}\label{Hamiltonian}
\mathcal{H}=-\frac{\kappa^{2}}{12a}\Pi_{a}^{2}+\frac{1}{2a^{3}}\Pi_{\varphi_{1}}^{2}
-\frac{1}{2a^{3}}\Pi_{\varphi_{2}}^{2}+\frac{a^{3}}{\kappa^{2}}\Lambda
+a^{3}V-\frac{3a}{\kappa^{2}}\mathcal{K}.
\end{align}
As is customary, the Hamiltonian is set to vanish, meaning $\mathcal{H} \equiv 0$. Utilizing (\ref{momenta}), one easily finds that this constraint is identical to Friedmann equation:
\begin{align}\label{eq5}
H^{2}=\frac{\kappa^{2}}{3}\rho +\frac{\Lambda}{3}-\frac{\mathcal{K}}{a^{2}}.
\end{align}
where \( H = \dot{a}/a \) represents the Hubble parameter, and \( \rho \) denotes the energy density of the quintom field, which is defined as:
\begin{align}\label{density}
\rho \equiv \frac{1}{2}\dot{\varphi}_{1}^{2}-\frac{1}{2}\dot{\varphi}_{2}^{2}
+V(\varphi_{1},\varphi_{2}).
\end{align}
When \( |\dot{\varphi}_{1}| > |\dot{\varphi}_{2}| \), representing the Canonical scalar Field Dominated (CFD) epoch, the configuration space lacks classically forbidden regions. This is due to the indefiniteness of the total kinetic term. In contrast, if \( |\dot{\varphi}_{1}| < |\dot{\varphi}_{2}| \), which corresponds to the Phantom scalar Field Dominated (PFD) regime, the negative definiteness of the total kinetic term ensures that only a specific region remains classically permitted:
\begin{align}\label{con1}
V(\varphi_{1},\varphi_{2})+\frac{\Lambda}{\kappa^{2}}-\frac{3}{\kappa^{2}}
\frac{\mathcal{K}}{a^{2}} \geq 0.
\end{align}
The ordinary and phantom scalar fields obey the Klein-Gordon equations, which are obtained by varying action (\ref{action2}) with respect to $\varphi_{1}$ and $\varphi_{2}$, respectively:
\begin{align}
\ddot{\varphi}_{1}+3H\dot{\varphi}_{1}+\frac{\partial V}{\partial \varphi_{1}}&=0,\label{phi1} \\
\ddot{\varphi}_{2}+3H\dot{\varphi}_{2}+\frac{\partial V}{\partial \varphi_{2}}&=0.\label{phi2}
\end{align}
Assuming a perfect-fluid energy-momentum tensor, the pressure of the quintom field can be defined as:
\begin{align}\label{pressure}
P \equiv \frac{1}{2}\dot{\varphi}_{1}^{2}-\frac{1}{2}\dot{\varphi}_{2}^{2}
-V(\varphi_{1},\varphi_{2}).
\end{align}
Now, it may easily be understood that equations (\ref{phi1})-(\ref{phi2}) are equivalent to the conservation equation:
$\dot{\rho}+3H(\rho +P)=0$.
Note that the conservation equation for cosmological constant $\Lambda$ is fulfilled through another equation of state which is expressed as:
$P_{\Lambda}=-\rho_{\Lambda}=-\Lambda /\kappa^{2}$.\\
By the use of (\ref{eq5}), the second-order equation for the scale factor is obtained:
\begin{align}\label{eq55}
\dot{H}+H^{2}=-\frac{\kappa^{2}}{6}\left( \rho +3P \right)+\frac{\Lambda}{3}.
\end{align}

By employing (\ref{density}) and (\ref{pressure}) and considering a constant barotropic index $\omega$, one can readily derive a relationship connecting the quintom potential to the quintom scalar field:
\begin{align}\label{12}
V(\varphi_{1},\varphi_{2})=\frac{1-\omega}{1+\omega}\left( \frac{1}{2}\dot{\varphi}_{1}^{2}-\frac{1}{2}\dot{\varphi}_{2}^{2} \right), \quad \omega \neq -1.
\end{align}
Hence, a similar result is found with the virial theorem, according to which the kinetic energy is proportional to the potential energy of the field.

\subsection{Classical quintom trajectory for no quintom potential\label{subsect.2.2}}
In this subsection, we aim to explore a simplified model where the cosmological constant is set to zero ($\Lambda = 0$) and the quintom potential vanishes, $V(\varphi_{1}, \varphi_{2}) = 0$. The associated equation of state, $\omega$, takes a value of $1$, corresponding to stiff matter characterized by $P = \rho$. This behavior stands in stark contrast to the current state of the universe. However, such an evolutionary scenario might be valid in ekpyrotic or cyclic cosmological models, where this type of matter could dominate during the collapsing phase of the universe's evolution~\cite{R44}.

For the PFD case, a negative energy density under these assumptions renders the model unsuitable for representing dark energy, which is conventionally associated with a positive energy density. Despite violating all known energy conditions, the model intriguingly exhibits certain phantom-like features and benefits from being mathematically straightforward. On the other hand, for the CFD case, while the energy density is positive and aligns with expectations for dark energy, the pressure remains problematic---it should be negative for dark energy but fails to meet this criterion, making this case similarly flawed. Consequently, this model serves merely as a theoretical exercise or ``toy model'' to examine underlying features. Notably, as we will show, this toy model demonstrates characteristics reminiscent of a ``big rip'' scenario in configuration space, making it an interesting exploratory framework. We will address more realistic models in subsequent subsections.

In order to acquire classical solutions within the quintom framework, we have to impose $\mathcal{K} = -1$ in the condition outlined in (\ref{con1}). If $\mathcal{K}$ is instead set to $0$, it becomes evident during calculations that at least one of the scalar fields would emerge as a purely imaginary function. This outcome leads to complications involving a double scalar field problem---phantom or ordinary---rather than a viable quintom model. In this work, we focus exclusively on configurations involving the quintom field.

One of the primary goals of this study is to construct wave packets using the Wheeler-DeWitt equation. For this purpose, it is critical to identify a classical trajectory within configuration space that eliminates classical time $t$. The necessity arises from the absence of such a time parameter in the Wheeler-DeWitt equation.

Owing to the cyclic nature of the variables $\varphi_{1}$ and $\varphi_{2}$, the corresponding momentum components $\Pi_{\varphi_{1}}$ and $\Pi_{\varphi_{2}}$ are conserved quantities. Utilizing equation (\ref{momenta}), we reach at:
\begin{align}
\dot{\varphi}_{1}^{2}=\frac{C_{1}^{2}}{a^{6}}, \qquad
\dot{\varphi}_{2}^{2}=\frac{C_{2}^{2}}{a^{6}},
\end{align}
where $C_{1}$ and $C_{2}$ are constants.
Consequently, based on equation (\ref{eq5}), the following result is obtained:
\begin{align}
\frac{\mathrm{d}\varphi_{j}}{\mathrm{d}a}=\pm \frac{|C_{j}|}{a
\sqrt{a^{4}+\frac{\kappa^{2}}{6}\left( C_{1}^{2}-C_{2}^{2} \right)}};
\qquad j=1,2,
\end{align}
which can be easily integrated to result in
\begin{align}\label{solphi}
\varphi_{j}=\mp \sqrt{\frac{3}{2}}\frac{|C_{j}|}{\kappa \sqrt{C_{12}}} \arctan \left( \frac{\sqrt{6a^{4}-\kappa^{2}C_{12}}}{\kappa \sqrt{C_{12}}} \right),
\end{align}
where \( C_{12} = C_1^2 - C_2^2 \), with the assumption that \( |C_1| \neq |C_2| \). This assumption is straightforward, as it not only prevents the denominator from equating to zero but also ensures that the phantom and canonical scalar fields do not neutralize each other.
It is important to note that in the PFD regime (where $C_{12} < 0$) and the CFD regime (where $C_{12} > 0$), after performing a few algebraic manipulations, equation (\ref{solphi}) can be respectively simplified to
\begin{align}\label{eq15}
\varphi_{j}&=\pm \sqrt{\frac{3}{2}}\frac{|C_{j}|}{\kappa \sqrt{|C_{12}|}}
\arccos \left( \frac{\kappa \sqrt{|C_{12}|}}{\sqrt{6}a^{2}} \right),\\ \label{eq16}
\varphi_{j}&=\pm \sqrt{\frac{3}{2}}\frac{|C_{j}|}{\kappa \sqrt{C_{12}}}
\, \mathrm{arcsinh} \left( \frac{\kappa \sqrt{C_{12}}}{\sqrt{6}a^{2}} \right).
\end{align}
The classical trajectory (eq. \ref{eq15}) has a minimum scale factor value given by
\linebreak
\( a_{\mathrm{min}} = \sqrt[4]{\kappa^{2} |C_{12}|/6} \) and extends to infinite scale factor values at finite scalar field values, \( \varphi_{j} = \pm \sqrt{\frac{3}{2}} \frac{|C_{j}|}{\kappa \sqrt{|C_{12}|}} \frac{\pi}{2} \). As such, it exhibits behaviors resembling a big rip solution. However, it is important to highlight that the scale factor diverges to infinity only as time approaches \( \pm \infty \). Additionally, the density, \( \rho \), scales proportionally to \( a^{-6} \), which corresponds to a stiff-fluid density scaling. While this toy model does not fully capture a conventional big rip scenario, it does share some notable characteristics of such behavior in configuration space, making it an intriguing subject of study.

\begin{figure*}[ht]
\centering
\includegraphics[width=\textwidth]{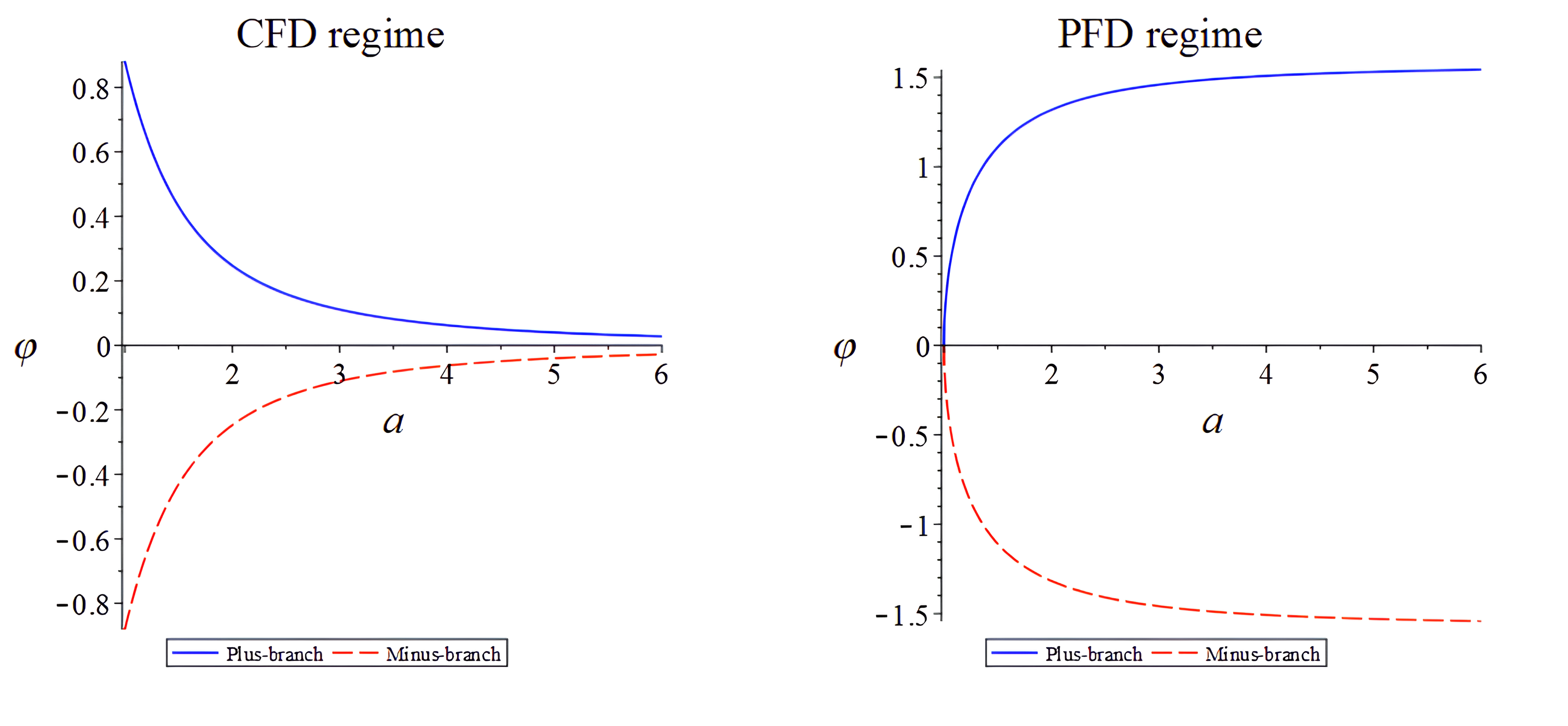}\\
\caption{This figure demonstrates the qualitative behaviors of the solutions of PFD regime, (\ref{eq15}), CFD regime, (\ref{eq16}). In plotting, we have set $\kappa = \sqrt{6}$, $C_{j}=2$, and $C_{12}=1$.}\label{fig1}
\end{figure*}

Solution (\ref{eq16}) does not contain any turning point; it just demonstrates two branches for which $a \to \infty$ if $\varphi_{j} \to 0$, and $a \to 0$ if
$\varphi_{j} \to \pm \infty$.\\
The qualitative behaviors of these solutions, eqs. (\ref{eq15})-(\ref{eq16}), are illustrated in Figure~\ref{fig1}. As is observed, in the PFD regime, the classical trajectory has a minimum scale factor value, \(a_{\mathrm{min.}}\), and progresses to an infinite scale factor at a finite scalar field value. This behavior is reminiscent of a big rip scenario. Notably, a bifurcation occurs at the start, with the scalar field increasing as time progresses. Conversely, in the CFD case, the classical trajectory lacks any turning points and instead consists of two distinct branches, where the scalar field decreases over time.
\subsection{Classical quintom trajectory for interacting exponential scalar field potential and vanishing cosmological constant\label{subsect.2.3}}
The selection of potential plays a pivotal role in our discussion. Hence, we set it as established in~\cite{q34}:
\begin{align}\label{exppot}
V(\varphi_{1},\varphi_{2})=V_{0}\,
e^{-\kappa(\lambda_{1}\varphi_{1}+\lambda_{2}\varphi_{2})},
\end{align}
where \( V_{0} \), \( \lambda_{1} \), and \( \lambda_{2} \) represent positive constants.
This type of potential has been thoroughly investigated in numerous assisted inflation studies; see, for instance, refs. \cite{ain1,ain2,ain3,ain4}. As demonstrated in \cite{q34}, the model under consideration serves as a counterexample to the typical behavior of quintom models with exponential potentials, as it permits the existence of either tracking attractors (with $\omega = 0$) or phantom attractors (where 
\linebreak
$\omega < -1$). Exponential potentials commonly arise in frameworks such as Kaluza-Klein theory, supergravity, superstring theory, and higher-order gravity.

In this subsection, we focus on a flat universe where $\mathcal{K}=0$. Based on equation (\ref{con1}), the quintom model does not possess a classically forbidden region. By transforming the classical equations of motion, (\ref{phi1})-(\ref{phi2}) and (\ref{eq55}), into a dynamical system constrained by the Friedmann equation (\ref{eq5}), it becomes evident that this system admits an attractor solution. This solution follows straightforward trajectories in configuration space, as outlined in~\cite{q34}:
\begin{align}\label{ctr01}
\varphi_{1}(\alpha)=\frac{\lambda_{1}}{\kappa}\alpha,
\qquad
\varphi_{2}(\alpha)=- \frac{\lambda_{2}}{\kappa}\alpha,
\end{align}
where $\alpha \equiv \ln (a)$. The condition for the existence of this attractor point is given by $\lambda^{2}_{1} - \lambda^{2}_{2} < 1$. Additionally, the point is stable if $\lambda_{1} < \sqrt{(1 + \lambda^{2}_{2}) / 2}$; otherwise, it becomes unstable. While the exact solutions were not explicitly derived in ref.~\cite{q34}, they can be straightforwardly determined as follows:
\begin{align}
\varphi_{1}&=\frac{2\lambda_{1}}{\left( \lambda^{2}_{1}-\lambda^{2}_{2} \right)\kappa} \ln \left[ 1+\frac{\left( \lambda^{2}_{1}-\lambda^{2}_{2} \right)H_{0}}{2}(t-t_{0}) \right],\\
\varphi_{2}&=\frac{-2\lambda_{2}}{\left( \lambda^{2}_{1}-\lambda^{2}_{2} \right)\kappa} \ln \left[ 1+\frac{\left( \lambda^{2}_{1}-\lambda^{2}_{2} \right)H_{0}}{2}(t-t_{0}) \right],\\
\frac{a}{a_{0}}&=\left[ 1+\frac{\left( \lambda^{2}_{1}-\lambda^{2}_{2} \right)H_{0}}{2}(t-t_{0}) \right]^{\frac{2}{\lambda^{2}_{1}-\lambda^{2}_{2}}},
\end{align}
where $H_{0}$, $a_{0}$, and $t_{0}$ are constants.
Utilizing (\ref{eq5}) and (\ref{density}) and expressing (\ref{eq5}) in the form\footnote{Nowhere in this paper does the comma index refer to differentiation, but merely to dependence.}
\begin{align*}
E_{\mathrm{kin.},\varphi_{1}}-E_{\mathrm{kin.},\varphi_{2}}
+E_{\mathrm{pot.}}=1,
\end{align*}
the components of kinetic energy and the resulting effective kinetic energy corresponding to the attractor solution mentioned above are respectively given by:
\begin{align*}
&E_{\mathrm{kin.},\varphi_{j}}=\frac{\kappa^{2}}{6}
\left(\frac{\mathrm{d}\varphi_{j}}{\mathrm{d}\alpha} \right)^{2}=
\frac{\lambda^{2}_{j}}{6};
\qquad j=1,2,\\
&E_{\mathrm{kin.,eff.}}
\equiv E_{\mathrm{kin.},\varphi_{1}}-E_{\mathrm{kin.},\varphi_{2}}
=\frac{\lambda^{2}_{1}-\lambda^{2}_{2}}{6}.
\end{align*}
As a result, they remain constant, which in turn means the potential energy of the quintom field is also unchanging:
\begin{align*}
E_{\mathrm{pot.}}=\frac{\kappa^{2}V}{3H^{2}}=
1-\frac{\lambda^{2}_{1}-\lambda^{2}_{2}}{6}
\end{align*}
It is useful to define the following sub-potentials:
\begin{align}\label{potsdef}
&E_{\mathrm{pot.,1}} \equiv \frac{1}{2}-\frac{\lambda^{2}_{1}}{6},
\qquad
E_{\mathrm{pot.,2}} \equiv \frac{1}{2}+\frac{\lambda^{2}_{2}}{6},\\
&\implies E_{\mathrm{pot.,eff.}} =E_{\mathrm{pot.,1}}+E_{\mathrm{pot.,2}}
\end{align}
The equation of state parameter of the quintom field would be:
\begin{align}
\omega=-1+\frac{1}{3}\left( \lambda^{2}_{1} -\lambda^{2}_{2} \right).
\end{align}
Therefore, for \(\lambda_{2} > \lambda_{1}\), the quintom field corresponds to a PFD regime, where \(\omega < -1\). In contrast, when \(\lambda_{2} < \lambda_{1}\) (the CFD regime), the scalar field behaves such that \(\omega > -1\).\\
The energy density of quintom field scales as
\begin{align*}
\rho = \rho_{0} \left(\frac{a}{a_{0}} \right)^{\lambda^{2}_{2}-\lambda^{2}_{1}},
\end{align*}
where $\rho_{0}$ is constant. Thus, this leads to a big-rip singularity for $\lambda_{2}>\lambda_{1}$ (PFD case) because in the limit
$t \to t_{1} \equiv t_{0} - 2/[(\lambda^{2}_{1}-\lambda^{2}_{2})H_{0}]$
both the energy density and the scale factor diverge. On the other hand, as \( t \to \infty \), both the scale factor and energy density approach zero. This behavior contrasts with the CFD case ($\lambda_{2} < \lambda_{1}$); In the limit \( t \to t_{1} \), the scale factor collapses to zero while the energy density diverges, yielding a big bang. However, for \( t \to \infty \), both the scale factor and energy density again diminish to zero.

\subsection{Classical quintom trajectory for scalar field fluid and negative cosmological constant\label{subsect.2.4}}
In cosmological models featuring a negative cosmological constant, it is possible to derive a straightforward set of classical solutions. Unlike a positive cosmological constant, which induces cosmological repulsion, a negative cosmological constant acts as a source of attraction. This attractive nature can counteract the effects of dark energy with negative pressure, such as that arising from cosmic strings or domain walls. Consequently, models incorporating a negative cosmological constant and specific fluid components can evolve symmetrically between two singularities, with an extremum occurring in between. As will be indicated later, it is even possible for the evolution to occur between two big rips within a finite cosmic time frame.

In this subsection, we consider a flat universe (denoted by $\mathcal{K} = 0$), a negative cosmological constant ($\Lambda < 0$), and a fluid characterized by the barotropic equation of state $P = \omega \rho$. By applying the energy conservation equation under these conditions, we arrive at:
\begin{align}
\rho = C_{3} a^{-3(1+\omega)},
\end{align}
where $C_{3}$ is constant. By the use of this equation, equations (\ref{eq5}) and (\ref{eq55}) can be readily solved to determine the scale factor:
\begin{align}\label{sinh}
a(t)=\left[a_{1}\sinh\left( \sqrt{\frac{ \Lambda}{3}}\, \beta t \right)  \right]^{1/ \beta},
\end{align}
where $a_{1}=\sqrt{\kappa^{2}C_{3}/ \Lambda}$ and $\beta=3(1+\omega )/2$ which is equivalent to
\begin{align}\label{sin}
a(t)=\left[a_{2}\sin \left( \sqrt{\frac{- \Lambda}{3}}\, |\beta | t \right)  \right]^{1/ \beta},
\end{align}
in which $a_{2}=\sqrt{\kappa^{2}C_{3}/ (-\Lambda)}$.
It is important to note that both `$\cosh$' and `$\cos$' are also valid solutions for our system. However, `$\sinh$' proves to be a better option compared to `$\cosh$', as `$\sinh$' represents both decelerated and accelerated eras, whereas $\cosh$ only captures the accelerating phase. Nonetheless, we continue with (\ref{sin}) due to our focus on a negative cosmological constant. By employing the definition of $\beta$, we can rewrite (\ref{12}) in the following form:
\begin{align}\label{formofV}
V(\varphi_{1}(t),\varphi_{2}(t))=\frac{3-\beta}{2\beta}\left( \dot{\varphi}_{1}^{2}-\dot{\varphi}_{2}^{2} \right)
\end{align}
which permits us to write (\ref{density}) down as
\begin{align}
\rho = \frac{3}{2\beta}\left( \dot{\varphi}_{1}^{2}-\dot{\varphi}_{2}^{2} \right)
=\frac{3}{\kappa^{2}}H^{2}-\frac{\Lambda}{\kappa^{2}}.
\end{align}
Therefore, we get a constraint equation for the evolution of $\varphi_{1}$ and $\varphi_{2}$:
\begin{align}
\dot{\varphi}_{1}^{2}-\dot{\varphi}_{2}^{2}=
-\frac{2\beta \Lambda}{3\kappa^{2}}\csc^{2} \left( \sqrt{\frac{- \Lambda}{3}}\, |\beta | t  \right)
\end{align}
If both the canonical and phantom fields are real, the equation represents a hyperbola, while if $\varphi_{1}$ is real and $\varphi_{2}$ is purely imaginary, the equation corresponds to a circle at a fixed cosmic time.
To analyze the system, it is essential to establish a relationship between the ordinary and phantom fields for their proper determination. Notably, in most studies (for instance, see refs. \cite{q34, behjcap}), $\varphi_{1}$ and $\varphi_{2}$ have been obtained proportional to each other. Based on this, we define $\varphi_{1} = A\varphi_{2}$, where $A$ is a constant. It is important to note that $A \neq 1$, as this condition prevents the two fields from canceling each other.
By adopting these assumptions, we can proceed to calculate the evolution of the scalar fields as follows:
\begin{align}\label{varphi1}
\varphi_{1}(t)&=\frac{\pm \sqrt{2} A}{\kappa \sqrt{\left|\beta \left( A^{2}-1 \right)\right|}}
\ln \left| \tan \left( \sqrt{\frac{- \Lambda}{12}}\, |\beta | t  \right) \right|,\\ \label{varphi2}
\varphi_{2}(t)&=\frac{\pm \sqrt{2}}{\kappa \sqrt{\left|\beta \left( A^{2}-1 \right)\right|}}
\ln \left| \tan \left( \sqrt{\frac{- \Lambda}{12}}\, |\beta | t  \right) \right|.
\end{align}
Obviously, the argument, i.e., $\sqrt{-\Lambda /12}\, |\beta| t$, must be between $0$ and $\pi$.  It is worth noting that when the argument becomes an integer multiple of $\pi$, a singularity arises.

\begin{figure*}[ht]
\centering
\includegraphics[width=\textwidth]{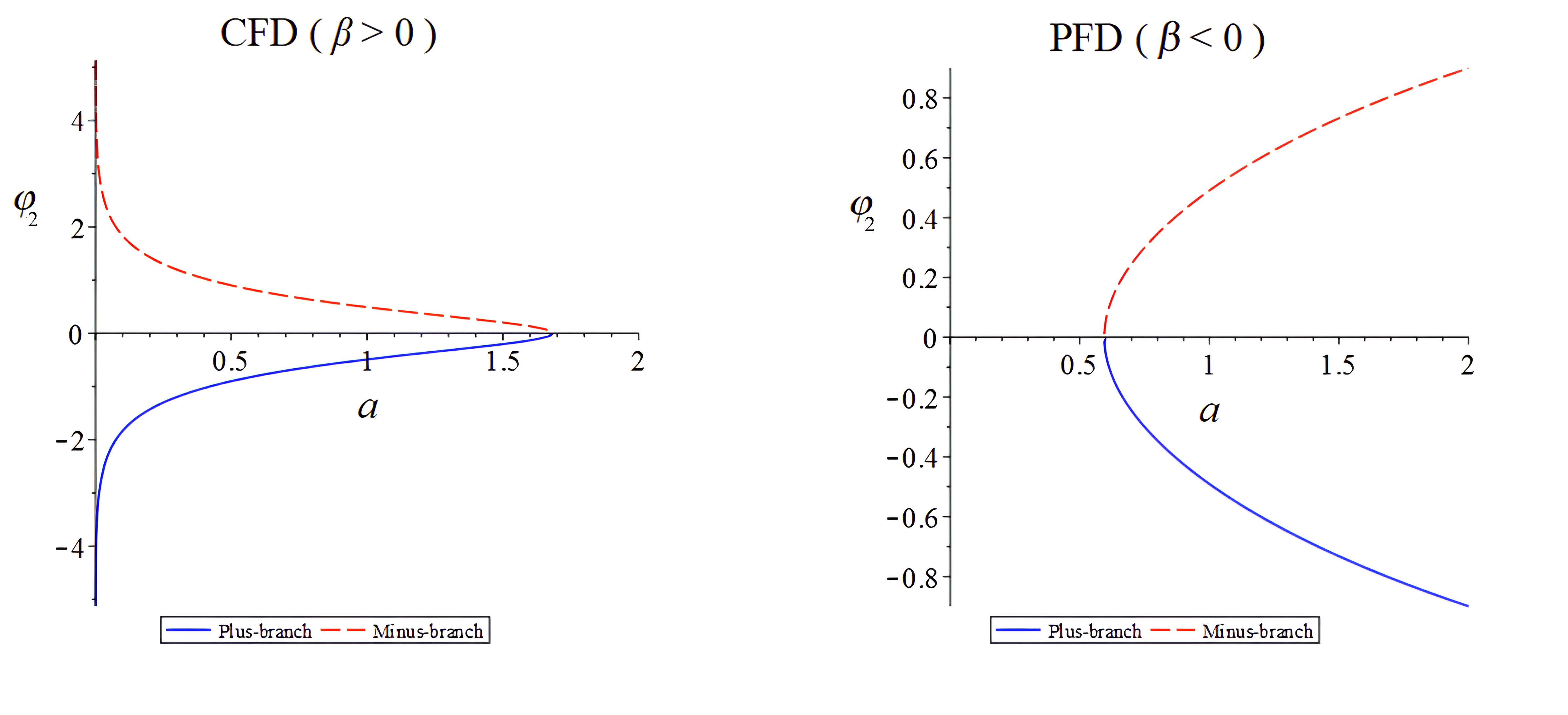}\\
\caption{This figure indicates the qualitative behavior of (\ref{cs42}) in two regime. In drawing these plots, we have chosen $\kappa = \sqrt{6}$, $|\beta|=2$, $A=2$, and $a_{2}=\sqrt{8}$.}\label{fig2}
\end{figure*}

Considering the fact that $\beta$ can be expressed as $\left( \lambda^{2}_{1}-\lambda^{2}_{2} \right)/2$, it follows that $\beta > 0$ corresponds to the CFD scenario, while $\beta < 0$ pertains to the PFD regime.
Pursuant to (\ref{sin}), for a positive $\beta$---which signifies a negative cosmological constant combined with an $\omega > -1$ fluid---the universe evolves from a big bang at $t=0$, reaches a maximum expansion characterized by $a_{\mathrm{max.}}=a_{2}^{1/\beta}$, and ultimately concludes with a big crunch at $t=\pi$. In contrast, in the PFD regime (negative $\beta$), the universe begins with a big rip at $t=0$, shrinks to a minimum scale factor $a_{\mathrm{min.}}=a_{2}^{-1/|\beta|}$, and expands toward another big rip at $t=\pi$. This evolution in the PFD case is symmetric, making it particularly intriguing from a theoretical perspective. Moreover, this framework may provide valuable insights into the study of the cosmological arrow of time. Interestingly, similar symmetric behavior can also be observed in the configuration space. By eliminating (\ref{sin}) and (\ref{varphi1})-(\ref{varphi2}) from the expressions for the classical time coordinate, the classical evolution trajectories are derived as follows:
\begin{align}\label{cs41}
\varphi_{1}(a)&=\frac{\pm \sqrt{2}A}{\kappa \sqrt{\left|\beta \left( A^{2}-1 \right)\right|}}
\ln \left( \frac{a^{\beta}}{a_{2}+\sqrt{a^{2}_{2}-a^{2\beta}}} \right),\\
\label{cs42}
\varphi_{2}(a)&=\frac{\pm \sqrt{2}}{\kappa \sqrt{\left|\beta \left( A^{2}-1 \right)\right|}}
\ln \left( \frac{a^{\beta}}{a_{2}+\sqrt{a^{2}_{2}-a^{2\beta}}} \right).
\end{align}
From these expressions, we observe the existence of two branches. In the PFD case, both branches extend indefinitely, meaning that as \( a \to \infty \), we find \( \varphi_{1,2} \to \pm \infty \). Each branch reaches a minimum, where \( \varphi_{1,2}(a) = 0 \), occurring at \( a_{\text{min}} = a_2^{-1/|\beta|} \). On the other hand, in the CFD regime, it is evident that the system features a maximum value at \( a_{\text{max}} \). The classical trajectories within the configuration space are illustrated in Figure~\ref{fig2}. According to this figure, in the case of the PFD regime, it is evident that both branches extend infinitely, indicating that as the scale factor grows without bound, the absolute value of the scalar field in both branches also diverges to infinity. Each branch exhibits a minimum point where the scalar field becomes zero, corresponding to a finite value of the scale factor. Initially, a bifurcation occurs, marking a split in the behavior of the scalar field. For the CFD regime, similarly, two branches are observed, but here the absolute value of the scalar field in both cases decreases over time, with the scale factor reaching a finite maximum value.

Using equations (\ref{varphi1})-(\ref{varphi2}) along with (\ref{formofV}), one can straightforwardly reconstruct the form of the quintom potential:
\begin{align}\label{quintompot}
V(\varphi_{1},\varphi_{2})=V_{01} \cosh^{2}\left( \frac{\varphi_{1}}{C_{01}} \right)+V_{02} \cosh^{2}\left( \frac{\varphi_{2}}{C_{02}}\right),
\end{align}
in which
\begin{align}
&C_{01}=\frac{\pm \sqrt{2} A}{\kappa \sqrt{\left|\beta \left( A^{2}-1 \right)\right|}},
\qquad
C_{02}=\frac{\pm \sqrt{2}}{\kappa \sqrt{\left|\beta \left( A^{2}-1 \right)\right|}},\\
&V_{01}=\frac{\beta \Lambda (\beta -3)}{6}C^{2}_{01}, \qquad V_{02}=-\, \frac{\beta \Lambda (\beta -3)}{6}C^{2}_{02}.
\end{align}
Various potential forms have been outlined in refs.~\cite{pot1,pot2}, according to which this specific type (\ref{quintompot}) is known as the unified dark matter potential.\\
It is worth noting that in the CFD case, the potential is positive only when $\beta < 3$ (i.e., $\omega < 1$).
\section{Quantum Quintom Cosmology\label{sect.3}}
This section focuses on exploring quantum behavior. For this purpose, the steps and conditions under examination align with those used in classical studies.
\subsection{Wheeler-DeWitt equation and quintom duality\label{subsect.3.1}}
As a result of quantizing the Hamiltonian constraint (\ref{Hamiltonian}), we can obtain the Wheeler-DeWitt equation. Choosing the Laplace-Beltrami factor ordering it turns out to be
\begin{align}\label{wdweqor}
\left(\frac{\kappa^{2}\hbar^{2}}{12}a\frac{\partial}{\partial a}a\frac{\partial}{\partial a}-\frac{\hbar^{2}}{2}\frac{\partial^{2}}{\partial \varphi^{2}_{1}}+\frac{\hbar^{2}}{2}\frac{\partial^{2}}{\partial \varphi^{2}_{2}}
+a^{6}\left( V(\varphi_{1},\varphi_{2})+\frac{\Lambda}{\kappa^{2}}\right)
-\frac{3\mathcal{K}a^{4}}{\kappa^{2}} \right) \psi (a,\varphi_{1},\varphi_{2})=0.
\end{align}
Similar to phantom duality which has been defined as~\cite{pd1,pd2}
\begin{align}
a \to \frac{1}{\bar{a}}, \qquad \varphi \to -i \bar{\varphi},
\end{align}
where $i=\sqrt{-1}$, we may define here quintom duality as follows:
\begin{align}
a \to \frac{1}{\bar{a}}, \qquad \varphi_{1} \to -i \bar{\varphi}_{1}, \qquad \varphi_{2} \to -i \bar{\varphi}_{2}.
\end{align}
Under the defined quintom duality, for flat universe, $\mathcal{K}=0$, the Wheeler-DeWitt equation for $a$, $\varphi_{1}$, and $\varphi_{2}$, i.e.,
\begin{align}\label{33}
\left(\frac{\kappa^{2}\hbar^{2}}{12}a\frac{\partial}{\partial a}a\frac{\partial}{\partial a}-\frac{\hbar^{2}}{2}\frac{\partial^{2}}{\partial \varphi^{2}_{1}}+\frac{\hbar^{2}}{2}\frac{\partial^{2}}{\partial \varphi^{2}_{2}}
+a^{6}\left( V(\varphi_{1},\varphi_{2})+\frac{\Lambda}{\kappa^{2}}\right)
\right) \psi (a,\varphi_{1},\varphi_{2})=0,
\end{align}
transforms into the Wheeler-DeWitt equation for $\bar{a}$, $\bar{\varphi}_{1}$, and $\bar{\varphi}_{2}$, viz.
\begin{align}
\left(\frac{\kappa^{2}\hbar^{2}}{12}\bar{a}\frac{\partial}{\partial \bar{a}}\bar{a}\frac{\partial}{\partial \bar{a}}+\frac{\hbar^{2}}{2}\frac{\partial^{2}}{\partial \bar{\varphi}^{2}_{1}}-\frac{\hbar^{2}}{2}\frac{\partial^{2}}{\partial \bar{\varphi}^{2}_{2}}
+\bar{a}^{6}\left( V(i\bar{\varphi}_{1},i\bar{\varphi}_{2})+\frac{\Lambda}{\kappa^{2}}\right)
\right) \psi (\bar{a},\bar{\varphi}_{1},\bar{\varphi}_{2})=0.
\end{align}
The transformations for both canonical and phantom fields are thus just a Wick rotation. We will revisit and discuss the results of quintom duality in subsection~\ref{subsect.3.4}.

The natural logarithm of scale factor, $\alpha =\ln (a)$, allows eq.~(\ref{33}) to be conveniently rewritten as follows
\begin{align}\label{331}
\left(\frac{\kappa^{2}\hbar^{2}}{12}\frac{\partial^{2}}{\partial \alpha^{2}}-\frac{\hbar^{2}}{2}\frac{\partial^{2}}{\partial \varphi^{2}_{1}}+\frac{\hbar^{2}}{2}\frac{\partial^{2}}{\partial \varphi^{2}_{2}}
+e^{6\alpha}\left( V(\varphi_{1},\varphi_{2})+\frac{\Lambda}{\kappa^{2}}\right)
\right) \Psi (\alpha ,\varphi_{1},\varphi_{2})=0,
\end{align}
We intend to utilize both forms (\ref{wdweqor}) and (\ref{331}) to work in the next subsections.

\subsection{Quantum quintom cosmology for no quintom potential\label{subsect.3.2}}
For vanishing cosmological constant and quintom potential and 
\linebreak
$\mathcal{K}=-1$, the Wheeler-DeWitt equation~(\ref{wdweqor}) may be obtained by the separation method, viz.
\begin{align}
\psi_{k_{1,2}}(a,\varphi_{1},\varphi_{2})=\Upsilon_{k_{1,2}}(a)\,
\phi_{k_{1}}(\varphi_{1})\,
\phi_{k_{2}}(\varphi_{2}).
\end{align}
We choose
\begin{align}
\phi_{k_{1}}(\varphi_{1})=e^{-ik_{1}\varphi_{1}/\hbar}, \qquad
\phi_{k_{2}}(\varphi_{2})=e^{-ik_{2}\varphi_{1}/\hbar},
\end{align}
because if real exponentials were used, wave functions for $\varphi_{1,2} \to \pm \infty$ would exponentially increase, which would not reflect classical behavior. By inserting these selections in (\ref{wdweqor}), one gets the following equation for $\Upsilon_{k_{1,2}}(a)$:
\begin{align}\label{aeq}
a^{2}\Upsilon_{k_{1,2}}^{\prime \prime}+a\Upsilon_{k_{1,2}}^{\prime}
+\frac{1}{\hbar^{2}}\left(\frac{36}{\kappa^{4}} a^{4}-\frac{6}{\kappa^{2}} k^{2}_{3} \right)\Upsilon_{k_{1,2}}=0,
\end{align}
where
$k^{2}_{3}=k^{2}_{2}-k^{2}_{1}$
and primes denote derivatives with respect to the scale factor $a$.
For convenience, let us set
$\kappa^{2}=6$.
We begin by examining the PFD case, characterized by $k_{2}>k_{1}$, which implies $k_{3}>0$. In general, the solutions to the governing equation are given by Bessel functions of the form $Z_{k_{3}/2\hbar}(a^{2}/2\hbar)$. However, to ensure the solution reflects the behavior of a classical trajectory, it is necessary to impose a boundary condition such that $\psi(a, \varphi_{1}, \varphi_{2}) \to 0$ as $a \to 0$. By applying this boundary condition, the path exhibits a minimum in configuration space relative to the scale factor, aligning with classical expectations. To satisfy this condition, we select the Bessel function $J_{k_{3}/2\hbar}(a^{2}/(2\hbar))$, which corresponds to $k_{3}>0$. The connection to the classical solution is achieved via the formal WKB approximation in the limit
\linebreak
$\hbar \to 0$. Consequently, it becomes necessary to analyze an asymptotic expansion of $J$ under the condition that both the argument and the index are large. According to ref.~\cite{abram1}, we utilize the following expression:
\begin{align}
J_{\nu}(\nu z)=\left( \frac{4\zeta}{1-z^{2}} \right)^{1/4}
\left( \frac{\mathrm{Ai}\left( \nu^{2/3}\zeta \right)}{\nu^{1/3}}
+\frac{\exp \left( -\frac{2}{3}\nu \zeta^{3/2} \right)}{1+\nu^{1/6} |\zeta|^{1/4}}\, \mathcal{O}\left( \frac{1}{\nu^{4/3}} \right)
 \right),
\end{align}
where $\mathrm{Ai}$ represents the Airy function, and the specific form of $\zeta$ depends on whether $z^2 \geq 1$ or $z^2 < 1$. Let us first analyze the case where $z^2 \geq 1$. Define $\nu = k_3 / (2\hbar)$ and $z = a^2 / k_3$. In this scenario (i.e., $z^2 \geq 1 \implies a^4 / k_3^2 \geq 1$), the expression for $\zeta$ is given by
\begin{align}
\zeta =- \left( \frac{3}{2}\left(\sqrt{\frac{a^{4}}{k_{3}^{2}}-1}- \arccos \left( \frac{k_{3}}{a^{2}} \right)\right)\right)^{2/3}.
\end{align}
Furthermore, it becomes essential to utilize the asymptotic expression for the Airy function, as its argument is also significantly large. Thus we make use~\cite{abram2}:
\begin{align}
\mathrm{Ai}\left( \left(\frac{k_{3}}{2\hbar} \right)^{2/3}\zeta \right)
\sim \pi^{-1/2} \left( -\left( \frac{k_{3}}{2\hbar} \right)^{2/3}\zeta \right)^{-1/4}\sin (\theta_{k_{3}}),
\end{align}
in which
\begin{align}
\theta_{k_{3}}=-\frac{k_{3}}{3\hbar}\zeta^{3/2}+\frac{\pi}{4}.
\end{align}
Through the principle of constructive interference, the classical trajectory can be recovered. The objective is to identify the point where the phase of the wave function reaches its extremum with respect to $k$. The phase is
\begin{align}
S_{k_{1,2}} \equiv \theta_{k_{3}}\pm \frac{k_{1}\varphi_{1}+k_{2}\varphi_{2}}{\hbar}.
\end{align}
The requirements
\begin{align}
\left.\frac{\partial S_{k_{1,2}}}{\partial k_{1}}\right|_{k_{1}=\bar{k}_{1}}=0, \qquad
\left.\frac{\partial S_{k_{1,2}}}{\partial k_{2}}\right|_{k_{2}=\bar{k}_{2}}=0,
\end{align}
yield (\ref{eq15}).
One may easily identify that $\bar{k}_{3}=\sqrt{|C_{12}|}$, in which $\bar{k}^{2}_{3}=\bar{k}^{2}_{2}-\bar{k}^{2}_{1}$, and $|C_{1}|=\bar{k}_{1}$ and $|C_{2}|=\bar{k}_{2}$. Note that we have set $\kappa^{2}=6$.

For the latter one, namely $z^{2}=(a^{4}/k^{2}_{3}) <1$, according to the expressions in ref.~\cite{abram2}, $\zeta <0$ and the corresponding Airy function decays exponentially. This result is not surprising because $z^{2}=(a^{4}/k^{2}_{3})<1$ corresponds to the classically forbidden region.

It can be easily indicated that through substitutions $\Pi_{a} \to \partial S_{k_{1,2}}/\partial a$, $\Pi_{\varphi_{1}} \to \partial S_{k_{1,2}}/\partial \varphi_{1}$, and $\Pi_{\varphi_{2}} \to \partial S_{k_{1,2}}/\partial \varphi_{2}$, $S_{k_{1,2}}$ is a solution to the Hamilton-Jacobi equation arising from~(\ref{Hamiltonian}).

Now we consider CFD case. Obviously, for this regime, the sign beside $k^{2}_{3}\text{-term}$ will change in (\ref{aeq}) meaning that we can rewrite it as $(\pm i k_{3})^{2}$ preserving the structure of (\ref{aeq}). Hence, the solutions to $\Upsilon_{k_{1,2}}(a)$ would be the Bessel functions $J_{\pm i k_{3}/(2\hbar)}(a^{2}/(2\hbar))$. Both solutions are permissible because no classically forbidden regions exist. Again, the classical trajectory in this case, i.e., (\ref{eq16}), like the previous one, is recovered through the principle of constructive interference.  As a consequence, one obtains two branches of (\ref{eq16}) from two Bessel functions. To suppress interference effects (and thereby avoid nonclassical behavior), it therefore is advisable to select one or the other Bessel function at a large scale factor. Given the hyperbolic nature of (\ref{wdweqor}) in the CFD regime, there is flexibility in imposing boundary conditions at constant scale factors. This choice includes imposing either one wave packet or two wave packets for each scalar field, depending on whether one prefers to demonstrate a single branch or both branches of the classical solution.

For the PFD case, however, the Wheeler-DeWitt equation becomes elliptic. Here, one can only apply boundary conditions such that \( \psi(a, \varphi_1, \varphi_2) \to 0 \) as \( a \to 0 \), while ensuring that it oscillates at most at other boundaries. This constraint leads to solutions expressed as \( J_{k_3 / (2\hbar)}\left(a^{2} / (2\hbar)\right) \) or superpositions thereof.
To construct a wave packet, one would explicitly consider the following superposition:
\begin{align}
\psi(a, \varphi_{1},\varphi_{2})=\int_{0}^{\infty}\int_{0}^{\infty}
\mathrm{d}k_{1} \mathrm{d}k_{2} A(k_{1})B(k_{2}) e^{-ik_{1}\varphi_{1}/\hbar}
e^{-ik_{12}\varphi_{2}/\hbar}
J_{k_{3}/(2\hbar)}(a^{2}/(2\hbar)),
\end{align}
where $A(k_{1})$ and $B(k_{2})$ are functions of $k_{1}$ and $k_{2}$ that are sharply localized around specific values \( \bar{k}_{1} \) and \( \bar{k}_{2} \), such as Gaussian profiles.  Since the phase of the Bessel function varies slowly with respect to \( k_{3} \), it follows that the wave packet would not exhibit significant dispersion near the classical trajectory's minimum—contrasting with the behavior observed in a massive scalar field, as described in ref.~\cite{q035}.
Nevertheless, dispersion phenomena are anticipated for larger scale factors. For a more practical and detailed examination of this scenario, the explicit analysis will be presented in the next subsection.

Drawing an analogy to ordinary quantum mechanics, the solution in the elliptic case can be likened to an initial wave function $\psi|_{t=0}$. On the other hand, the hyperbolic case corresponds to the time evolution of the wave function, as equation (\ref{wdweqor}) would feature a distinct set of foliations associated with an intrinsic time defined by $a$. This intrinsic time could potentially serve as a physical time, within which other degrees of freedom could evolve dynamically.
\subsection{Quantum quintom cosmology for interacting exponential potential and vanishing cosmological constant\label{subsect.3.3}}
In subsection \ref{subsect.2.3}, the quintom model was analyzed for a nonzero interacting exponential potential in a flat universe. Here, we extend the study to its quantum counterpart.
To proceed systematically, and without loss of generality, we first assume that the wave function can be expressed as $\Psi(\alpha, \varphi_{1}, \varphi_{2}) = \Psi_{1}(\alpha, \varphi_{1}) \Psi_{2}(\alpha, \varphi_{2})$ in equation (\ref{331}). Under this assumption, the problem reduces to two coupled partial differential equations:
\begin{align}\label{ceq1}
&\left(\hbar^{2}\frac{\partial^{2}}{\partial \alpha^{2}}-\hbar^{2}\frac{\partial^{2}}{\partial \varphi^{2}_{1}}
+e^{6\alpha}V(\varphi_{1},\varphi_{2})
\right) \Psi_{1} (\alpha ,\varphi_{1})=0,\\ \label{ceq2}
&\left(\hbar^{2}\frac{\partial^{2}}{\partial \alpha^{2}}+\hbar^{2}\frac{\partial^{2}}{\partial \varphi^{2}_{2}}
+e^{6\alpha}V(\varphi_{1},\varphi_{2})
\right) \Psi_{2} (\alpha ,\varphi_{2})=0.
\end{align}
Since equation (\ref{ceq1}) primarily addresses the Canonical field and equation (\ref{ceq2}) emphasizes the Phantom field, we refer to these as the C-part and P-part of the Wheeler-DeWitt equations, respectively.
These parts of the Wheeler-DeWitt equation can be effectively solved by implementing transformations to new variables, allowing the effective potential in front of $\Psi_{1,2}$ in (\ref{ceq1}) and (\ref{ceq2}) to be eliminated. This becomes achievable by initially transitioning to the coordinate sets $(Q_{1},Q_{2})$ and $(Q_{3},Q_{4})$, which are defined as follows:
\begin{align}
\text{For (\ref{ceq1})}:&\; \left\{ \begin{array}{l}
Q_{1} \equiv \alpha +\varphi_{1} \\
Q_{2} \equiv \alpha -\varphi_{1}
\end{array} \right. ; \\
\text{For (\ref{ceq2})}:&\; \left\{ \begin{array}{l}
Q_{3} \equiv \alpha +i \varphi_{2} \\
Q_{4} \equiv \alpha -i \varphi_{2}
\end{array} \right. .
\end{align}
These coordinates exhibit a light-cone type structure. It is worth noting that, in this mapping, a degree of freedom is initially added following the transformation. However, due to the relation \( Q_{1} + Q_{2} = Q_{3} + Q_{4} \), the system effectively reduces back to three degrees of freedom. Consequently, eqs.~(\ref{ceq1}) and (\ref{ceq2}) simplify to
\begin{align*}
&\left(\hbar^{2}\frac{\partial^{2}}{\partial Q_{1} \partial Q_{2}}+F(Q_{1},Q_{2},Q_{3},Q_{4})
\right) \Psi_{1} (Q_{1},Q_{2})=0,\\
&\left(\hbar^{2}\frac{\partial^{2}}{\partial Q_{3} \partial Q_{4}}+F(Q_{1},Q_{2},Q_{3},Q_{4})
\right) \Psi_{2} (Q_{3},Q_{4})=0,
\end{align*}
where $F(Q_{1},Q_{2},Q_{3},Q_{4})$ for the potential (\ref{exppot}) would be
\begin{align*}
\frac{1}{2}V_{0}\exp \left(3(Q_{1}+Q_{2})-\sqrt{\frac{3}{2}}
\left[\lambda_{1}(Q_{1}-Q_{2})-i\lambda_{2}(Q_{3}-Q_{4})\right]\right)
\end{align*}
Clearly, we require transformations into new variables that can effectively eliminate \( F \). These transformations are respectively represented as follows:
\begin{align}
\label{array1}
&\left\{\begin{array}{l}
u_{1}(\alpha,\varphi_{1},\varphi_{2})=
2\sqrt{V_{0}}\; \frac{\exp\left[3\alpha -\sqrt{6}(\lambda_{1}\varphi_{1}
+\lambda_{2}\varphi_{2})/2\right]}{6-\lambda^{2}_{1}}
\left( \cosh(X_{1})+\frac{\lambda_{1}}{\sqrt{6}}\sinh(X_{1}) \right);
\\
v_{1}(\alpha,\varphi_{1},\varphi_{2})=
2\sqrt{V_{0}}\; \frac{\exp\left[3\alpha -\sqrt{6}(\lambda_{1}\varphi_{1}
+\lambda_{2}\varphi_{2})/2\right]}{6-\lambda^{2}_{1}}
\left( \sinh(X_{1})+\frac{\lambda_{1}}{\sqrt{6}}\cosh(X_{1}) \right);
\end{array}
\right.\\ \label{array2}
&\left\{\begin{array}{l}
u_{2}(\alpha,\varphi_{1},\varphi_{2})=
2\sqrt{V_{0}}\; \frac{\exp\left[3\alpha -\sqrt{6}(\lambda_{1}\varphi_{1}
+\lambda_{2}\varphi_{2})/2\right]}{6+\lambda^{2}_{2}}
\left( \cosh(X_{2})-i\frac{\lambda_{1}}{\sqrt{6}}\sinh(X_{2}) \right);
\\
v_{2}(\alpha,\varphi_{1},\varphi_{2})=
2\sqrt{V_{0}}\; \frac{\exp\left[3\alpha -\sqrt{6}(\lambda_{1}\varphi_{1}
+\lambda_{2}\varphi_{2})/2\right]}{6+\lambda^{2}_{2}}
\left( -i \sinh(X_{2})-\frac{\lambda_{1}}{\sqrt{6}}\cosh(X_{2}) \right);
\end{array}
\right.
\end{align}
where
\begin{align}
X_{1}\equiv 3\varphi_{1}-\sqrt{\frac{3}{2}}\, \lambda_{1}\alpha,
\qquad
X_{2}\equiv i \left(3\varphi_{2}+\sqrt{\frac{3}{2}}\, \lambda_{2}\alpha \right).
\end{align}
As a result, the C- and P-parts of the Wheeler-DeWitt equation in these new coordinates turn out to be
\begin{align}\label{re1}
&\hbar^{2} \left( \frac{\partial^{2} \Psi_{1}}{\partial u^{2}_{1}}-
\frac{\partial^{2} \Psi_{1}}{\partial v^{2}_{1}} \right)+\Psi_{1}=0;\\ \label{re2}
&\hbar^{2} \left( \frac{\partial^{2} \Psi_{2}}{\partial u^{2}_{2}}+
\frac{\partial^{2} \Psi_{2}}{\partial v^{2}_{2}} \right)+\Psi_{2}=0.
\end{align}
For C-part, the new variables $u_{1}$ and $v_{1}$ transform the Wheeler-DeWitt equation into a Klein-Gordon-like form (\ref{re1}), where $u_{1}$ and $v_{1}$ act as ``time-like'' and ``space-like'' coordinates in minisuperspace, respectively. This separation isolates the dynamics of the canonical field $\varphi_{1}$, simplifying the analysis of wave-packet solutions. For P-part, the transformed equation becomes (\ref{re2}) resembling a Helmholtz equation. Here, $u_{2}$ and $v_{2}$ mix the phantom field $\varphi_{2}$ and the logarithm of the scale factor $\alpha$, capturing the phantom field's negative kinetic energy. The imaginary term in $X_{2}$ reflects the ghost-like nature of $\varphi_{2}$.\\
Utilizing WKB-approximations ansatz, $\Psi_{1}=C_{4} \exp (\pm i S_{1}/\hbar)$ and \linebreak
$\Psi_{2}=C_{5} \exp (\pm i S_{2}/\hbar)$, we may acquire at lowest order the Hamiltonian-Jacobi equations
\begin{align}\label{74.1}
&\left( \frac{\partial S_{01}}{\partial u_{1}} \right)^{2}-
\left( \frac{\partial S_{01}}{\partial v_{1}} \right)^{2}=1,\\ \label{74.2}
&\left( \frac{\partial S_{02}}{\partial u_{2}} \right)^{2}+
\left( \frac{\partial S_{02}}{\partial v_{2}} \right)^{2}=1.
\end{align}
These equations may be solved through a separation ansatz by
\begin{align}\label{csr}
S_{01,k_{1}}=k_{1}u_{1}-\sqrt{k^{2}_{1}-1}\,v_{1}, \qquad
S_{02,k_{2}}=k_{2}u_{2}-i \sqrt{k^{2}_{2}-1}\,v_{2}.
\end{align}
It is also feasible to solve the Hamilton-Jacobi equation by employing actions with differing signs applied to the pairs \(\{u_{1}, v_{1}\}\) and \(\{u_{2}, v_{2}\}\). These are obtained from the one chosen above through rotations in the $(u_{1},v_{1})$- and $(u_{2},v_{2})$-planes. In doing so, all solutions may be mapped onto each other for (\ref{74.2}) because of the rotational symmetry inherent in this equation. If we set $\lambda_{1}<6$, then only two solutions of the C-part may be mapped onto each other.

By the use of the classical actions $S_{01,k_{1}}$ and $S_{02,k_{2}}$, the equations of motion take the form:
\begin{align*}
\left.\frac{\partial S_{01,k_{1}}}{\partial k_{1}}\right|_{k_{1}=\bar{k}_{1}}=C_{6}, \qquad
\left.\frac{\partial S_{02,k_{2}}}{\partial k_{2}}\right|_{k_{2}=\bar{k}_{2}}=C_{7}.
\end{align*}
The classical trajectories (\ref{ctr01}) are recovered through the following special selections, respectively:
\begin{align*}
&\bar{k}^{2}_{1}=E^{-1}_{\mathrm{pot.,1}}=\left(\frac{1}{2}
-\frac{\lambda^{2}_{1}}{6}  \right)^{-1}, \qquad C_{6}=0;\\
&\bar{k}^{2}_{2}=E^{-1}_{\mathrm{pot.,2}}=\left(\frac{1}{2}
+\frac{\lambda^{2}_{2}}{6}  \right)^{-1}, \qquad C_{7}=0,
\end{align*}
namely we get
\begin{align}
\varphi_{1}(\alpha)=\frac{\lambda_{1}}{\sqrt{6}}\alpha,
\qquad
\varphi_{2}(\alpha)=-\, \frac{\lambda_{2}}{\sqrt{6}}\alpha.
\end{align}

Incorporating this lowest-order ansatzs into the Wheeler-DeWitt equations, it can be determined that the equations are already fully satisfied. Consequently, the corresponding exact wave packets for the Wheeler-DeWitt equations are derived as follows:
\begin{align}
\Psi_{1}(u_{1},v_{1})=\int \mathrm{d}k_{1} \, A(k_{1})\biggl(
&C_{8}\exp \left[ \frac{i}{\hbar}\left( k_{1}u_{1}-\sqrt{k^{2}_{1}-1}\, v_{1} \right) \right]\nonumber \\ & +
C_{9}\exp \left[ -\, \frac{i}{\hbar}\left( k_{1}u_{1}-\sqrt{k^{2}_{1}-1}\, v_{1} \right) \right]
\biggl),\\
\Psi_{2}(u_{2},v_{2})=\int \mathrm{d}k_{2} \, B(k_{2})\biggl(
&C_{10}\exp \left[ \frac{i}{\hbar}\left( k_{2}u_{2}-i\sqrt{k^{2}_{2}-1}\, v_{2} \right) \right]\nonumber \\ & +
C_{11}\exp \left[ -\, \frac{i}{\hbar}\left( k_{2}u_{2}-i \sqrt{k^{2}_{2}-1}\, v_{2} \right) \right]
\biggl),
\end{align}
and the total $\Psi$ is given by $\Psi_{1}\Psi_{2}$.
Through constructive interference, these equations can be used to recover classical trajectories. We opt for the amplitudes $A$ and $B$ Gaussian functions with width $\sigma$ centered around $\bar{k}_{1}$ and $\bar{k}_{2}$, respectively:
\begin{align*}
A(k_{1})&=\frac{1}{\sqrt[4]{\pi \sigma^{2} \hbar^{2}}} \exp \left[-\, \frac{\left( k_{1}-\bar{k}_{1} \right)^{2}}{2\sigma^{2} \hbar^{2}}  \right],\\
B(k_{2})&=\frac{1}{\sqrt[4]{\pi \sigma^{2} \hbar^{2}}} \exp \left[-\, \frac{\left( k_{2}-\bar{k}_{2} \right)^{2}}{2\sigma^{2} \hbar^{2}}  \right].
\end{align*}
Choosing $C_{8}=C_{9}$ and $C_{10}=C_{11}$ for definiteness, we arrive at wave packets of the forms
\begin{align}\label{wp1}
\psi_{1}(u_{1},v_{1}) & \approx C_{8}\, \pi^{1/4}\, \sqrt{\frac{2\sigma \hbar}{1-i\sigma^{2}\hbar S^{\imath \imath }_{01,k_{1}}}}\,
\exp \left[ \frac{i S_{01,k_{1}}}{\hbar}-\frac{\sigma^{2} S^{\imath 2}_{01,k_{1}}}{2\left( 1-i\hbar \sigma^{2}S^{\imath \imath}_{01,k_{1}} \right)}  \right],\\ \label{wp2}
\psi_{2}(u_{2},v_{2}) & \approx C_{10}\, \pi^{1/4}\, \sqrt{\frac{2\sigma \hbar}{1-i\sigma^{2}\hbar S^{\imath \imath}_{02,k_{2}}}}\,
\exp \left[ \frac{i S_{02,k_{2}}}{\hbar}-\frac{\sigma^{2} S^{\imath 2}_{02,k_{2}}}{2\left( 1-i\hbar \sigma^{2}S^{\imath \imath}_{02,k_{2}} \right)}  \right],
\end{align}
where the Taylor expansions of $S_{01,k_{1}}$ and $S_{02,k_{2}}$ have been performed around $\bar{k}_{1}$ and $\bar{k}_{2}$, respectively, and for simplicity $S_{0j,k_{j}}(\bar{k}_{j})\equiv S_{0j,k_{j}}$ for $j=1,2$. The imath symbols, ($^{\imath}$), for $S_{01,k_{1}}$ and $S_{02,k_{2}}$ denote derivatives with respect to $k_{1}$ and $k_{2}$, respectively. It must be mentioned that the terms of the orders $(k_{1}-\bar{k}_{1})^{3}$ and $(k_{2}-\bar{k}_{2})^{3}$ have been neglected in the exponents. This simplification is justified when Gaussian functions are sharply peaked around \(\bar{k}_{1}\) and \(\bar{k}_{2}\), which occurs if the parameter \(\sigma\) is sufficiently small. Due to the fact that $S^{\imath}_{01,k_{1}}(\bar{k_{1}})=0$ and $S^{\imath}_{02,k_{2}}(\bar{k_{2}})=0$ define the classical trajectories, the packets are peaked around them. Consequently, both wave packets adhere to the classical trajectories while spreading as \( v^{2}_{1} \to \infty \) and \( v^{2}_{2} \to \infty \). This behavior is evident from the expressions in
\linebreak
(\ref{wp1})-(\ref{wp2}); the Gaussian widths increase indefinitely due to contributions from terms proportional to
\(
[S^{\imath\imath}_{01,k_{1}}(\bar{k}_{1})]^2
\)
and
\(
[S^{\imath\imath}_{02,k_{2}}(\bar{k}_{2})]^2,
\)
as described by the following:
\begin{align}
S^{\imath \imath}_{01,k_{1}}(\bar{k}_{1})
=\frac{v_{1}}{\left( \bar{k}^{2}_{1}-1 \right)^{3/2}},
\qquad
S^{\imath \imath}_{02,k_{2}}(\bar{k}_{2})
=\frac{i v_{2}}{\left( \bar{k}^{2}_{2}-1 \right)^{3/2}}.
\end{align}
The mentioned fact may also be readily recognized from the absolute square of the wave packets
\begin{align}
\left| \psi_{1}(u_{1},v_{1}) \right|^{2} & \approx |C_{8}|^{2} \, \sqrt{\pi}\,
\frac{2\sigma \hbar}{\sqrt{1+\sigma^{4} \hbar^{2} (S^{\imath \imath}_{01,k_{1}})^{2}}}\exp \left[-\, \frac{\sigma^{2} S^{\imath 2}_{01,k_{1}}}{1+\sigma^{4}\hbar^{2} (S^{\imath \imath}_{01,k_{1}})^{2}}  \right],\\
\left| \psi_{2}(u_{2},v_{2}) \right|^{2} & \approx |C_{10}|^{2} \, \sqrt{\pi}\,
\frac{2\sigma \hbar}{\sqrt{1+\sigma^{4} \hbar^{2} (S^{\imath \imath}_{02,k_{2}})^{2}}}\exp \left[-\, \frac{\sigma^{2} S^{\imath 2}_{02,k_{2}}}{1+\sigma^{4}\hbar^{2} (S^{\imath \imath}_{02,k_{2}})^{2}}  \right],
\end{align}
where we have neglected the complex conjugate parts in (\ref{wp1})-(\ref{wp2}).
Consequently, due to the nontrivial dispersion relation---or, more precisely, the nonlinear dependence of $S_{01,k_{1}}$ and $S_{02,k_{2}}$ on $\bar{k}_{1}$ and $\bar{k}_{2}$, respectively---spreading effects arise. It follows that the semiclassical approximation is invalid throughout configuration space.

Let us analyze and discuss each sub-state (C- and P-parts) individually.

In approaching to big-rip singularity, we have $u_{2} \to -\infty$ and $v_{2} \to -\infty$. Thus, this singularity dwells in a legitimate quantum region. Due to the fact that
\begin{align*}
v^{2}_{2} \sim \exp \left[6\alpha-\sqrt{6}(\lambda_{1}\varphi_{1}+\lambda_{2}\varphi_{2}) \right]
\equiv e^{6\alpha}V(\varphi_{1},\varphi_{2}),
\end{align*}
the nontrivial potential causes the dispersion.

As a result, the big-rip singularity is smoothed out---we are no longer able to use an approximate time parameter when wave packets disperse. This marks the end of time and classical evolution, leaving only a stationary quantum state. Such an outcome arises due to the influence of quantum gravity effects at extremely large scales.

Two inequivalent actions will exist for $u_{1}>0$ (i.e., when $\lambda_{1}<\sqrt{6}$). The construction of the wave packets is then based on
\begin{align*}
S_{01,k_{1}}&=k_{1}u_{1}-\sqrt{k^{2}_{1}-1}\, v_{1},\\
S_{01,k_{1}}&=-k_{1}u_{1}-\sqrt{k^{2}_{1}-1}\, v_{1}.
\end{align*}
Furthermore, the entire $(\alpha, \varphi_{1})\text{-plane}$ is mapped into a single quarter of the \((u_1, v_1)\text{-plane}\). However, only one solution meets this requirement---the trivial solution. In order to obtain a nontrivial solution, the boundary condition must be relaxed so that \(\Psi_1 = 0\) is enforced solely at the origin of the \((u_1, v_1)\text{-plane}\). On the lines $u_{1}=0$ and $v_{1}=0$, the wave packet does not vanish because of the non-normalizability of the wave packet in both $\alpha$ and $\varphi_{1}$. This stems from the fact that the associated classical trajectory lacks a turning point.

\begin{figure}
\centering
\begin{minipage}[c]{\textwidth}
\centering
    \includegraphics[width=\textwidth]{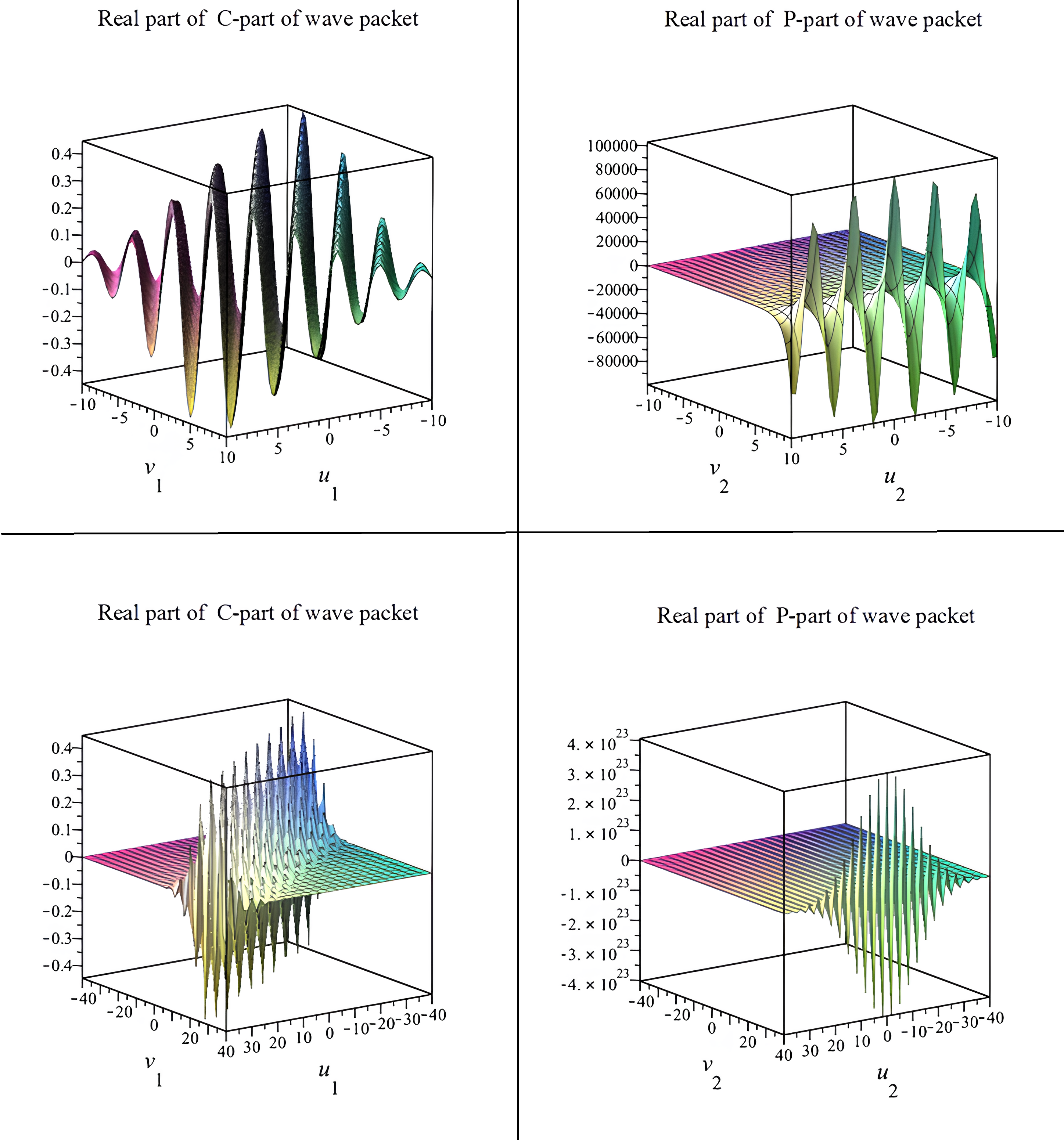}\\
\caption{This figure demonstrates the real part of both C- and P-parts of the wave packet in two different ranges for each part. In plotting, we have selected $\lambda_{1}=\sqrt{6}/4$, $\lambda_{2}=\sqrt{6}/5$, $\sigma = 0.1$, $\hbar = 1$, $C_{8}=\pi^{-1/4}$, and $C_{10}=2 \pi^{-1/4}$.}\label{fig3}
\end{minipage}
\end{figure}

By adopting this adjusted boundary condition, the wave packet vanishes at the big-bang singularity (i.e., \(\Psi_1 \to 0\) as \(\alpha \to -\infty\)) and spreads out as \(\alpha\) becomes large. This implies that quantum theory avoids the inclusion of a big-bang singularity. It is important to note that since the entire \((u_2, v_2)\text{-plane}\) corresponds to the entire \((\alpha, \varphi_2)\text{-plane}\), no analogous restriction applies to \(\Psi_2\). Visual representations of both parts of the wave packet are depicted in Figure~\ref{fig3} in two different ranges. As is observed, the wave packet for the P-part spreads near the classical singularity. For the C-part, one has $\psi_{1} \to 0$ at the origin. In each sector corresponding to one copy of the $(\alpha, \varphi)$-plane, the same wave packet propagates.
\subsection{Quantum quintom cosmology for scalar field fluid and negative cosmological constant\label{subsect.3.4}}
In the model analyzed in subsection~\ref{subsect.2.4}, the classical solutions necessitate a potential structured in the form
\begin{align*}
V(\varphi_{1},\varphi_{2})=V_{01} \cosh^{2}\left( \frac{\varphi_{1}}{C_{01}} \right)+V_{02} \cosh^{2}\left( \frac{\varphi_{2}}{C_{02}}\right)
\end{align*}
The Wheeler-DeWitt equation, therefore, takes the form
\begin{align}
\biggl[&\frac{\hbar^{2}}{2}\left(
\frac{\partial^{2}}{\partial \alpha^{2}}
-\frac{\partial^{2}}{\partial \varphi^{2}_{1}}
+\frac{\partial^{2}}{\partial \varphi^{2}_{2}}
\right) \nonumber \\ & 
+e^{6\alpha}\left(
V_{01} \cosh^{2}\left( \frac{\varphi_{1}}{C_{01}} \right)+V_{02} \cosh^{2}\left( \frac{\varphi_{2}}{C_{02}}\right)
+\frac{\Lambda}{6}
\right)
\biggl]\Psi(\alpha , \varphi_{1} , \varphi_{2})=0.
\end{align}
Classical singularities are located in a domain where the scalar fields $|\varphi_{1}|$ and $|\varphi_{2}|$ become large.
Therefore, when analyzing quantum behavior in this region, it is adequate to approximate the potential assuming large scalar field values
\begin{align}
\tilde{V}(\varphi_{1},\varphi_{2}) \approx
\frac{V_{01}}{4}e^{\pm 2 \varphi_{1}/|C_{01}|}
+\frac{V_{02}}{4}e^{\pm 2 \varphi_{2}/|C_{02}|},
\end{align}
where in the following the upper and lower signs refer to positive and negative scalar fields, respectively. Here, to facilitate the solution process, without loss of generality, we assume $\Psi(\alpha , \varphi_{1} , \varphi_{2})=\Psi_{1}(\alpha , \varphi_{1})\Psi_{2}(\alpha , \varphi_{2})$.
This approach renders the problem somewhat analogous to the scenario discussed in the previous subsection (see \ref{subsect.3.3}). Once again, it becomes necessary to apply certain transformations to the variables:
\begin{align}
&\left\{\begin{array}{l}
u_{1}(\alpha,\varphi_{1})=
3\sqrt{\frac{V_{01}}{2}}\; \frac{C^{2}_{01}\exp\left[3\alpha \pm(\varphi_{1}
/|C_{01}|)\right]}{9C^{2}_{01}-1}
\left( \cosh(X_{1})\mp \frac{1}{3|C_{01}|}\sinh(X_{1}) \right);
\\
v_{1}(\alpha,\varphi_{1})=
3\sqrt{\frac{V_{01}}{2}}\; \frac{C^{2}_{01}\exp\left[3\alpha \mp (\varphi_{1}
/|C_{01}|)\right]}{9C^{2}_{01}-1}
\left( \sinh(X_{1})\pm \frac{1}{3|C_{01}|}\cosh(X_{1}) \right);
\end{array}
\right.\\
&\left\{\begin{array}{l}
u_{2}(\alpha,\varphi_{2})=
3\sqrt{\frac{V_{02}}{2}}\; \frac{C^{2}_{02}\exp\left[3\alpha \pm(\varphi_{2}
/|C_{02}|)\right]}{9C^{2}_{02}+1}
\left( \cosh(X_{2})\pm \frac{i}{3|C_{02}|}\sinh(X_{2}) \right);
\\
v_{2}(\alpha,\varphi_{2})=
3\sqrt{\frac{V_{02}}{2}}\; \frac{C^{2}_{02}\exp\left[3\alpha \mp (\varphi_{2}
/|C_{02}|)\right]}{9C^{2}_{02}+1}
\left(-i \sinh(X_{2})\pm \frac{1}{3|C_{02}|}\cosh(X_{2}) \right);
\end{array}
\right.
\end{align}
where
\begin{align}
X_{1}\equiv 3\varphi_{1}\pm \frac{\alpha}{|C_{01}|},
\qquad
X_{2}\equiv i \left( 3\varphi_{2}\mp \frac{\alpha}{|C_{02}|} \right).
\end{align}
In these variables, the equations (\ref{re1})-(\ref{re2}) are recovered:
\begin{align}
&\hbar^{2} \left( \frac{\partial^{2} \Psi_{1}}{\partial u^{2}_{1}}-
\frac{\partial^{2} \Psi_{1}}{\partial v^{2}_{1}} \right)+\Psi_{1}=0;\\
&\hbar^{2} \left( \frac{\partial^{2} \Psi_{2}}{\partial u^{2}_{2}}+
\frac{\partial^{2} \Psi_{2}}{\partial v^{2}_{2}} \right)+\Psi_{2}=0.
\end{align}
The physical interpretations of the new variables $u$ and $v$ are the same as in the previous case, (\ref{array1})-(\ref{array2}):
The variables $u_{1}$ and $v_{1}$ again reduce the C-part to a Klein-Gordon form, isolating $\varphi_{1}$'s dynamics. By the use of $u_{2}$ and $v_{2}$, the P-part's equation becomes elliptic, reflecting the phantom field's instability. The variables $u_{2}$ and $v_{2}$ encode the interplay between $\varphi_{2}$ and $\alpha$ near singularities.\\
Starting from a WKB ansatz, one can derive solutions. The Hamilton-Jacobi equations are, once again, expressed through (\ref{74.1})-(\ref{74.2}). It is important to recognize that these equivalences are merely formal, as $u$ and $v$ variables were defined differently. These equations can still be effectively solved using (\ref{csr}), with the considerations discussed in subsection~\ref{subsect.3.3} about the selection of the action being applicable here as well. The equations of motion for
\begin{align*}
\left.\frac{\partial S_{01,k_{1}}}{\partial k_{1}}\right|_{k_{1}=\bar{k}_{1}}=0, \qquad
\left.\frac{\partial S_{02,k_{2}}}{\partial k_{2}}\right|_{k_{2}=\bar{k}_{2}}=0,
\end{align*}
are given by
\begin{align}
\varphi_{1,\pm}(\alpha)=\mp A \sqrt{\frac{|\beta|}{3|A^{2}-1|}} \, \alpha +c_{\bar{k}_{1}},\\
\varphi_{2,\pm}(\alpha)=\pm \sqrt{\frac{|\beta|}{3|A^{2}-1|}} \, \alpha +c_{\bar{k}_{2}},
\end{align}
where $c_{\bar{k}}$s are constants. These solutions are approximately aligned with the classical solutions (eqs. (\ref{cs41})-(\ref{cs42})). In what follows, we discuss the underlying reasons for this correspondence. First, it is essential to divide our analysis into two parts: CFD and PFD. This distinction arises because, in the former one $\beta$ is positive, whereas in the latter one it is negative. It is important to note that, due to quintom duality, the behavior of the scalar field differs between the two regimes. For the PFD case, a large scalar field corresponds to a large scalar field, while for CFD one, a large scalar field maps to a small scalar field.\\
Now, focusing on the PFD regime, if the classical solutions (eqs. (\ref{cs41})-(\ref{cs42})) are approximated for a large scale factor, we find that
\begin{align}
\varphi_{1,\pm}(\alpha)=\pm A \sqrt{\frac{|\beta|}{3|A^{2}-1|}} \, \alpha \pm |C_{01}| \ln (2a_{2}),\\
\varphi_{2,\pm}(\alpha)=\pm \sqrt{\frac{|\beta|}{3|A^{2}-1|}} \, \alpha \pm |C_{02}| \ln (2a_{2}),
\end{align}
where $\pm$ indicate the distinct branches of the solutions. Given that $a$ is large, it follows that $\alpha \geq 0$ in this scenario. Clearly, the limit corresponding to large positive scalar fields lies on the plus-branch, whereas for large negative scalar fields, it is obtained on the minus-branch.\\
In the CFD regime, if equations (\ref{cs41})-(\ref{cs42}) are approximated for a small scale factor, we arrive at
\begin{align}
\varphi_{1,\pm}(\alpha)=\pm A \sqrt{\frac{\beta}{3|A^{2}-1|}} \, \alpha \mp |C_{01}| \ln (2a_{2}),\\
\varphi_{2,\pm}(\alpha)=\pm \sqrt{\frac{\beta}{3|A^{2}-1|}} \, \alpha \mp |C_{02}| \ln (2a_{2}).
\end{align}
In this scenario, $\alpha \leq 0$ due to the smallness of $a$. As a result, the limit of large positive scalar fields are achieved on the minus-branches, and for large negative scalar fields on the plus-branches.\\
Therefore, our approximated solutions coincide with classical solutions.

By dint of the classical actions $S_{01,k_{1}}$ and $S_{02,k_{2}}$, the C- and P-parts of Wheeler-DeWitt equation can be solved. These equations are precisely satisfied through the WKB ansatz. The resulting wave packets take a similar form to those discussed in the previous subsection; however, this time they involve redefined $u$s and $v$s, along with alternative choices for the centers of the Gaussian functions, $\bar{k}_{1}$ and $\bar{k}_{2}$.  Just as in the case with vanishing $\Lambda$, the sub-wave packets spread as $v^{2}_{1} \to \infty$ and $v^{2}_{2} \to \infty$. The big-rip singularity emerges at $v^{2}_{2} \to \infty$ and $u_{2} \to \infty$. Again, the big-bang singularity does therefore not exist in the quantum theory. Consequently, the singularity remains again hidden in the quantum domain, thereby rendering the semiclassical approximation inapplicable throughout the configuration space. Furthermore, at big-bang, $\Psi \to 0$.

\section{Interesting and challenging questions\label{sect.4}}
Before terminating the paper, let us discuss some important questions especially for the reason that this paper treats the phantom field on the same footing as a normal scalar field, both classically and quantum mechanically. However, the quantum instability of phantom field is a major concern in the literature~\cite{qqq1,qqq2}.
\subsection{Is the Wheeler-DeWitt quantization consistent in the presence of negative kinetic energy terms?}

The Wheeler-DeWitt quantization in cosmological models, particularly those involving phantom fields (e.g., the quintom model), remains an active area of research in quantum cosmology. A key challenge arises from the presence of negative kinetic energy terms, which complicates both the classical and quantum descriptions of such systems.

\subsubsection{Challenges posed by negative kinetic terms}
Negative kinetic energy terms appear in two primary contexts:
\begin{itemize}
    \item \textit{The conformal mode of gravity}: In the Arnowitt-Deser-Misner (ADM) decomposition, the scale factor (or its logarithm) has a negative-definite kinetic term in the Einstein-Hilbert action~\cite{0f1,0f2}.
    \item \textit{Phantom (ghost) fields}: Scalar fields with negative kinetic terms contribute negatively to the Hamiltonian~\cite{q06}. Our study focuses on this case.
\end{itemize}

These terms introduce several difficulties in quantum cosmology:
\begin{enumerate}
    \item \textit{Instability and Unbounded Hamiltonian}: Negative kinetic energy implies the Hamiltonian is unbounded from below, leading to instabilities in the quantum theory, a well-known issue in ghost field theories~\cite{0f4}.
    \item \textit{Non-Normalizable Wave Functions}: In minisuperspace models, wrong-sign kinetic terms can cause wave functions to grow exponentially in certain directions, rendering them non-normalizable under the standard $L^2$ inner product~\cite{q030}.
    \item \textit{Problems with the Semiclassical Limit}: The Wheeler-DeWitt equation is expected to recover classical general relativity in the WKB limit. Negative kinetic terms disrupt this by introducing unstable modes that prevent a smooth classical limit~\cite{0f6}.
\end{enumerate}

\subsubsection{Quantum cosmological implications}
In the standard canonical quantization approach, negative kinetic terms lead to non-Hermitian contributions to the Hamiltonian, raising concerns about unitarity and stability. This is particularly problematic in quintom models, where the phantom field's dominance may induce ghost-like instabilities~\cite{0r1}.

The path integral formulation also faces challenges, as negative kinetic terms can cause divergent integrals or non-convergence due to non-positive-definite actions. Recent studies suggest that introducing suitable counterterms or redefining the Hamiltonian (e.g., through regularization of non-Hermitian contributions) might mitigate these issues~\cite{0r1}. For models combining phantom and quintessence fields, modifications to the potential or a redefinition of the action may be necessary to preserve quantization consistency.

Furthermore, the violation of standard energy conditions in phantom models complicates conservation laws and symmetries, affecting the interpretation of the Wheeler-DeWitt equation. Classically, negative kinetic energy often leads to big-rip singularities. While quantum effects might soften these singularities, the underlying instabilities persist, suggesting the quantum universe may inherit similar pathologies.

\subsubsection{Potential resolutions}
Several approaches have been proposed to address these issues:
\begin{enumerate}
    \item \textit{Klein-Gordon-Type Inner Product}: The Wheeler-DeWitt equation resembles a Klein-Gordon equation on superspace, allowing for a conserved inner product analogous to the Klein-Gordon case~\cite{0f7}. However, this does not fully resolve negative probabilities.
    \item \textit{Euclidean Path Integral (Wick Rotation)}: Hartle and Hawking's no-boundary proposal involves rotating the conformal factor to imaginary values, rendering the kinetic term positive~\cite{0f8}.
    \item \textit{York Time and Gauge Fixing}: Treating the conformal mode as a ``time'' variable (York time) can reinterpret the negative kinetic term, yielding a well-posed Schr\"{o}dinger-like equation~\cite{0f9}.
    \item \textit{Quantum Gravity Corrections}: Approaches like loop quantum cosmology modify the Wheeler-DeWitt equation to avoid negative kinetic energy problems near singularities~\cite{0f10}.
    \item \textit{Counterterms and Action Redefinitions}: As noted earlier, introducing counterterms or redefining the action (e.g., via auxiliary fields or non-canonical kinetic terms) can restore consistency in specific regimes~\cite{0r1}.
\end{enumerate}

Thus, the consistency of Wheeler-DeWitt quantization with negative kinetic terms depends on the interpretation and additional assumptions:
\begin{itemize}
    \item \textit{Without modifications}, the Wheeler-DeWitt equation is generally inconsistent due to instabilities and non-normalizability.
    \item \textit{With modifications} (e.g., Wick rotation, alternative inner products, counterterms, or quantum gravity corrections), certain formulations can retain consistency in specific regimes.
\end{itemize}

Therefore, in summary, the Wheeler-DeWitt quantization with negative kinetic terms is not manifestly consistent in its original form. However, strategies like counterterms, action redefinitions, or Hamiltonian regularization can restore consistency in certain cases. The issue remains open and is deeply connected to fundamental questions about time and quantization in gravity.
\subsection{How vacuum instability or unbounded Hamiltonians might affect the interpretation of the wave function of the universe?}
In the context of quantum cosmology, particularly within the quintom model, vacuum instability and unbounded Hamiltonians present significant challenges to interpreting the wave function of the universe. In three main parts we discuss this issue:
\begin{enumerate}
  \item \textit{Vacuum Instability in Quintom Models}: The quintom model introduces a phantom field with negative kinetic energy, which can lead to:
\begin{itemize}
  \item \textit{Unbounded Hamiltonian}: The phantom component’s energy density is not bounded from below, leading to instabilities in the quantum vacuum~\cite{q42}.
  \item \textit{Quantum instabilities}: The phantom field’s negative energy states can cause runaway particle production, affecting the semiclassical approximation used in quantum cosmology~\cite{ins2}.
\end{itemize}
These instabilities challenge the standard interpretation of the wave function of the universe (Hartle-Hawking or Vilenkin wave functions), as they imply that the background space-time may not remain stable under quantum fluctuations.
  \item \textit{Implications for the Wave Function of the Universe}: In quantum cosmology, the wave function is typically interpreted via the Wheeler-DeWitt equation. However:
      \begin{itemize}
        \item \textit{Unboundedness from below}: If the Hamiltonian is unbounded, the wave function may not admit a normalizable solution, making probabilistic interpretations problematic~\cite{ins3}.
        \item \textit{Tunneling vs. no-boundary proposal}:
        \begin{itemize}
          \item The Vilenkin tunneling proposal might favor an unstable vacuum by allowing transitions to arbitrarily negative energy states.
          \item The Hartle-Hawking no-boundary proposal relies on Euclidean path integrals, which may diverge if the action is unbounded ~\cite{ins4}.
        \end{itemize}
      \end{itemize}
  \item  \textit{Possible Resolutions and Alternative Interpretations}:
      \begin{itemize}
        \item \textit{Regularization via higher-derivative terms}: Some studies suggest modifying the phantom Lagrangian to stabilize the Hamiltonian~\cite{q40}.
        \item \textit{Non-standard quantization}: Alternative quantization schemes (e.g., polymer quantization) might provide a way to handle unbounded Hamiltonians~\cite{0f10}.
        \item \textit{Multiverse interpretation}: If the wave function describes a landscape of vacua, the instability may lead to transitions between different cosmological phases~\cite{ins7}.
      \end{itemize}
\end{enumerate}
Therefore, in a nutshell, vacuum instability and unbounded Hamiltonians in quintom models raise serious challenges for interpreting the wave function of the universe, particularly in terms of normalizability, tunneling probabilities, and the validity of semiclassical approximations. Resolving these issues may require modifications to the theory or alternative quantization approaches.

\subsection{What is the Hilbert space or inner product?}
  In this paper we did not use inner product of wave functions, but we just used real part of the probability density. Hence, we do not so sensitive in this regard, but let us discuss in this issue in brief.\\
  The wave function of the universe is given by
  \begin{align*}
  \psi(a,\varphi_{1},\varphi_{2})=\psi_{1}(a,\varphi_{1})\psi_{2}(a,\varphi_{2}).
  \end{align*}
  Since the total wave function is a product of two functions, each defined over different variables, the total Hilbert space is the tensor product of the individual Hilbert spaces. The Hilbert space $\mathcal{H}$ is supposed to be the space of square-integrable wave functions\footnote{The scale factor $a$ is restricted to $a>0$.}:
\begin{align*}
\mathcal{H}_{\mathrm{phantom}}=L^{2}(\mathbb{R}_{+}\times \mathbb{R},da \, d\varphi_{1}),\\
\mathcal{H}_{\mathrm{quintessence}}=L^{2}(\mathbb{R}_{+}\times \mathbb{R},da \, d\varphi_{2}).
\end{align*}
Then the Hilbert space of the total system would then be:
\begin{align*}
\mathcal{H}=\mathcal{H}_{\mathrm{phantom}} \otimes \mathcal{H}_{\mathrm{quintessence}}
=L^{2}(\mathbb{R}_{+}\times \mathbb{R}^{2},da \, d\varphi_{1}\, d\varphi_{2})
\end{align*}
The inner product on this Hilbert space for two total wave functions $\Psi$ and $\Phi$ is given by:
\begin{align*}
\langle \Phi | \Psi \rangle =\int_{0}^{\infty} \int_{-\infty}^{\infty} \int_{-\infty}^{\infty} \Phi^{\star}(a,\varphi_{1}, \varphi_{2})\Psi(a,\varphi_{1}, \varphi_{2}) da \, d\varphi_{1} \, d\varphi_{2}.
\end{align*}
If the wave function factors as
$\Psi(a,\varphi_{1},\varphi_{2})=\psi_{1}(a,\varphi_{1})\psi_{2}(a,\varphi_{2})$,
then under separability assumptions (if $\psi_{1}$ and $\psi_{2}$ do not interfere with each other via $a$) we might write:
\begin{align*}
\langle \Phi | \Psi \rangle =
\left(\int da \, d\varphi_{1}\, \phi_{1}^{\star}(a,\varphi_{1})\psi_{1}(a,\varphi_{1})  \right) \cdot \left( \int da \, d\varphi_{2}\, \phi_{2}^{\star}(a,\varphi_{2})\psi_{2}(a,\varphi_{2}) \right)
\end{align*}
The measures we used in defining Hilbert space and inner product were $d\mu_{1}=da \, d\varphi_{1}$ and $d\mu_{2}=da \, d\varphi_{2}$. Note that the choice of measure depends on the quantization scheme. In quantum cosmology, the measure is often determined by the DeWitt metric on minisuperspace. For a FRW universe with scalar fields, a typical choice is:
\begin{align*}
d\mu_{1} = \sqrt{-g}\, da \, d\varphi_{1}, \qquad d\mu_{2}=\sqrt{-g}\, da \, d\varphi_{2},
\end{align*}
where $\sqrt{-g}$ is the square root of the determinant of the minisuperspace metric. For a flat minisuperspace metric, this might reduce to:
\begin{align*}
d\mu_{1} = a^{n}\, da \, d\varphi_{1}, \qquad d\mu_{2}=a^{n}\, da \, d\varphi_{2},
\end{align*}
where $n$ depends on the chosen factor ordering (commonly $n=1$ or $n=0$ for simplicity).\\
The above mentioned approach would be our method if we were using inner product in this paper. But there is a problem here that we need to address a bit:\\
Because of the phantom field, which introduces a negative kinetic energy, the Hamiltonian becomes indefinite and may not be bounded from below. This leads to a non-positive definite inner product, such as:
\begin{align*}
\langle \Phi | \Psi \rangle = \int d\varphi_{1}\, d\varphi_{2}\, da\,
\mu (a, \varphi_{1}, \varphi_{2})\, \Phi^{\star}(a, \varphi_{1}, \varphi_{2}) \Psi(a, \varphi_{1}, \varphi_{2}).
\end{align*}
But due to the phantom field, this inner product can become indefinite, and the Hilbert space becomes a Krein space or a pseudo-Hilbert space, where not all norms are positive.\\
Consequences of an indefinite inner product can be listed as follows:
  \begin{enumerate}
    \item \textit{Ghost instabilities}: States with negative norm (from the phantom field) can grow unbounded, leading to unphysical predictions like negative probabilities.
    \item \textit{Breakdown of unitary evolution}: In quantum mechanics, unitary evolution preserves the norm of the wave function. In an indefinite metric space, this may no longer be guaranteed.
    \item \textit{Problematic vacuum structure}: No well-defined vacuum may exist, which makes particle interpretation and energy spectrum ambiguous.
  \end{enumerate}
There are several attempts in the literature to resolve the issue including:
  \begin{enumerate}
    \item \textit{Modified inner products}: Researchers attempt to define a new, conserved, positive-definite inner product using conditions from the Wheeler–DeWitt equation (e.g., following the Klein–Gordon inner product analogy).
    \item \textit{PT-symmetric or pseudo-Hermitian quantum mechanics}: Some studies (e.g., ref.~\cite{PT1}) have explored how pseudo-Hermitian Hamiltonians with indefinite kinetic terms can still define a consistent quantum theory.
    \item \textit{Alternative quantization schemes}: Such as loop quantum cosmology or using path integral approaches that circumvent the need for a canonical Hilbert space structure.
  \end{enumerate}
In a nutshell, in the quintom model, the Hilbert space structure is ill-defined in the usual sense due to the phantom field's negative kinetic term. The inner product is not positive-definite, leading to issues with unitarity and stability. A consistent quantum interpretation may require using pseudo-Hermitian quantum mechanics, Krein spaces, or alternative quantization frameworks to handle the indefinite metric structure. To show how it this plays out mathematically, we give the followings:\\
Because the Wheeler–DeWitt equation is hyperbolic (due to the minus sign in the kinetic term of phantom) we adopt a Klein-Gordon-like inner product:
  \begin{align*}
  \langle\Phi | \Psi \rangle = i \int d\varphi_{1}\, d\varphi_{2}\, a^{n}
  \left(\Phi^{\star} \frac{\partial \Psi}{\partial a}-\frac{\partial \Phi^{\star}}{\partial a}\Psi  \right),
  \end{align*}
  where $n$  can be chosen (e.g., $n=0$, $n=3$) to ensure covariance or finiteness. This is indefinite, meaning $\langle \Psi | \Psi \rangle$  can be negative or zero even for non-zero $\Psi$---reflecting the presence of ghost degrees of freedom. Its consequences are:
  \begin{enumerate}
    \item \textit{Negative norm states}: from the phantom field, we get states with negative norm---problematic for probability interpretation.
    \item \textit{Non-unitary evolution}: no guarantee of conservation of total probability.
    \item \textit{Hilbert space structure}: becomes a Krein space---a vector space with an indefinite inner product that splits into positive- and negative-norm subspaces.
  \end{enumerate}
  If one attempts with Schr\"{o}dinger-like inner product (positive-definite attempt), then he must define:
  \begin{align*}
  \langle \Phi | \Psi \rangle =\int da\, d\varphi_{1}\, d\varphi_{2}\,
  |\Phi^{\star}(a, \varphi_{1},\varphi_{2}) \Psi(a, \varphi_{1},\varphi_{2})|.
  \end{align*}
  But this does not conserve norm under Wheeler-DeWitt evolution (not a true Hilbert space). Again, ghosts obstruct standard probabilistic quantum mechanics.

  In the quintom loop quantum cosmology case, the Hamiltonian constraint becomes something like:
  $\hat{\mathcal{C}}\Psi (\nu , \varphi_{1},\varphi_{2})=0$, where $\nu$
  is related to the volume (discrete), and the evolution in $\nu$ is governed by recursion relation. The discreteness of geometry in loop quantum cosmology naturally regulates the exponential instabilities caused by phantom fields. Loop quantum cosmology offers a bounded, well-defined Hilbert space, where the inner product is defined via group averaging techniques:
  \begin{align*}
  \langle \Phi | \Psi \rangle = \sum_{\nu}^{}\int d\varphi_{1} \, d\varphi_{2} \, \overline{\Phi(\nu, \varphi_{1},\varphi_{2})}\, \Psi(\nu, \varphi_{1},\varphi_{2}).
  \end{align*}
  This can be made positive-definite in the loop quantum cosmology framework.
\subsection{ What is the interpretation of the wave function?}
  In quantum cosmology, the wave function plays a central role in describing the quantum state of the entire universe. Unlike in ordinary quantum mechanics, where the wave function describes the probabilistic behavior of particles, the wave function in quantum cosmology encodes the probability amplitudes for different configurations of the universe itself, including its geometry and matter content. Here are key interpretations:
  \begin{enumerate}
    \item \textit{Hartle-Hawking (No-Boundary) Proposal}: The wave function of the universe is interpreted as a path integral over compact Euclidean geometries (imaginary time) with no initial boundary, giving a quantum amplitude for the universe to transition from ``nothing'' to a classical configuration~\cite{0w1}.
    \item \textit{Vilenkin's Tunneling Proposal}: The wave function represents a universe tunneling ``from nothing'' (a quantum vacuum) into an expanding classical universe via a quantum tunneling process~\cite{0w2}.
    \item \textit{Bohmian Interpretation in Quantum Cosmology}: The wave function guides the deterministic evolution of the universe's geometry and matter fields, with a ``pilot wave'' acting on the configuration space of the universe~\cite{0w3}.
    \item \textit{Decoherence and Classicality}: The wave function predicts a superposition of universes, but decoherence selects branches where classical spacetime emerges (consistent with observations)~\cite{CK1}.
    \item \textit{Problem of Interpretation}: Unlike standard quantum mechanics, there is no external observer, so interpretations rely on intrinsic probabilities (e.g., the ``Born rule'' applied to cosmological histories)~\cite{0f1}.
  \end{enumerate}
  Therefore, in a nutshell, the wave function in quantum cosmology provides a quantum-mechanical description of the universe's initial conditions and possible histories, with interpretations varying based on boundary conditions (e.g., no-boundary vs. tunneling proposals) and philosophical stances (e.g., Everettian, Bohmian).
\subsection{Is the Wheeler-DeWitt equation treated as a physical evolution equation or a constraint?}
      The Wheeler-DeWitt equation is fundamentally treated as a constraint equation rather than a conventional physical evolution equation in quantum gravity and cosmology. Below are the key interpretations:
      \begin{enumerate}
        \item \textit{Constraint in Canonical Quantum Gravity}: \textit{i}. The Wheeler-DeWitt equation arises from the Hamiltonian formulation of general relativity, where it represents a quantum version of the classical Hamiltonian constraint (a component of the Einstein field equations). \textit{ii}. It enforces time reparametrization invariance (reflecting the diffeomorphism symmetry of general relativity), meaning it does not describe evolution in an external time but rather a static condition on the wave function of the universe~\cite{0f1}.
        \item \textit{Lack of External Time (Problem of Time)}: The WDW equation does not contain an explicit time parameter, leading to the ``problem of time'' in quantum cosmology. Time must emerge internally (e.g., via semiclassical approximations or relational observables)~\cite{0f7}.
        \item \textit{Attempts to Recover Evolution}: Some approaches (e.g., semiclassical approximations) extract an effective time parameter by treating part of the system classically (e.g., the scale factor as a ``clock'')~\cite{0c3}.
        \item \textit{Alternative Views (Timeless Formalism)}: In the relational quantum mechanics framework, the Wheeler-DeWitt equation is seen as a timeless constraint, and physical predictions are made via correlations between variables (e.g., matter fields as clocks)~\cite{0c4}.
        \item \textit{Criticism and Alternatives}: Some argue that the Wheeler-DeWitt equation is incomplete due to its neglect of quantum fluctuations beyond minisuperspace models. Proposals like loop quantum cosmology modify or replace it~\cite{0f10}.
      \end{enumerate}
      Thus, the Wheeler-DeWitt equation is primarily a constraint reflecting the diffeomorphism invariance of general relativity, though efforts to extract ``evolution'' rely on auxiliary structures (e.g., matter clocks, semiclassical approximations).
\subsection{What boundary conditions are physically justified in quantum cosmology (apart from vanishing the wave function when the scale factor is zero)?}
    I want to answer this question in general, not just for our specific model, to make it comparable. In quantum cosmology, boundary conditions for the Wheeler-DeWitt equation are crucial for selecting physically meaningful solutions and determining the quantum behavior of the universe. Beyond the simple requirement that the wave function $\psi$ vanishes when the scale factor approaches to zero $a \to 0$ (avoiding singularities), several physically motivated boundary conditions have been proposed. Below are the most prominent ones, along with their justifications and references:
    \begin{enumerate}
      \item \textit{No-Boundary Proposal (Hartle-Hawking)}~\cite{0w1}:\\
      The Hartle-Hawking no-boundary proposal suggests that the wave function of the universe is defined by a Euclidean path integral over compact, smooth four-dimensional geometries with no initial boundary (``no past''). This means the universe has no singular beginning but instead emerges from a quantum gravitational state where time is imaginary (Euclidean) near the origin and transitions smoothly to a classical Lorentzian space-time. The wave function takes the form $\psi \sim e^{-S_{\mathrm{E}}}$, where $S_{\mathrm{E}}$ is the Euclidean action, favoring universes that are regular and symmetric at small scales. While this proposal avoids the big-bang singularity, it does not naturally predict inflation unless additional assumptions are introduced.

      \item \textit{Tunneling Proposal (Vilenkin)}~\cite{0w2}:\\
      In contrast to the no-boundary condition, Vilenkin's tunneling proposal describes the universe as emerging from ``nothing'' (a quantum vacuum) via quantum tunneling into an expanding Lorentzian space-time. The wave function is constructed from outgoing modes only ($\psi_{\mathrm{V}} \sim e^{+i S_{L}}$), representing an expanding universe. This approach is analogous to quantum tunneling in particle physics and assigns higher probability to initial conditions that lead to inflation. However, the tunneling condition requires a definition of the ``nothing'' state, which remains conceptually challenging.

      \item  \textit{DeWitt's Singularity Avoidance Condition}~\cite{0f1}:\\
      DeWitt proposed that the wave function must vanish at the big-bang singularity ($\psi \to 0$ as $a \to 0$), ensuring that the universe avoids a classical singularity in the quantum regime. This condition is simple and physically motivated but does not uniquely determine the wave function without additional constraints (e.g., regularity).

      \item \textit{Robin (Mixed) Boundary Conditions}~\cite{00w1}:\\
      Robin boundary conditions generalize the singularity avoidance approach by allowing a combination of Dirichlet ($\psi =0$) and Neumann ($\partial \psi / \partial a =0$) conditions at $a \to 0$. This flexibility permits a broader class of wave functions (singularity-avoiding solutions), such as $\psi \sim \sin (ka)$ near $a=0$, which can be useful in minisuperspace models.

      \item  \textit{Loop Quantum Cosmology (LQC) Corrections}~\cite{0f10}:\\
      Loop quantum cosmology modifies the Wheeler-DeWitt equation near the Planck scale, replacing the classical singularity with a quantum bounce. The wave function remains finite at $a = 0$, describing a universe that contracts to a minimal volume before re-expanding. This approach is derived from loop quantum gravity and provides a concrete mechanism for singularity resolution.

      \item \textit{Path Integral Uniqueness (Linde's Proposal)}~\cite{0s1}:\\
      Linde suggested a modification of the Hartle-Hawking approach by allowing complex metrics (not strictly Euclidean or Lorentzian) in the gravitational path integral. This approach uses Wick rotation to define the path integral. This generalization can better accommodate inflationary initial conditions while retaining the conceptual framework of quantum creation.
    \end{enumerate}

      Table~\ref{table1} demonstrates a comparison of boundary conditions.
      \begin{table}[ht]
\caption{Comparison of Boundary Conditions}
\centering
\begin{tabular}{c c c c}
\hline\hline
Proposal & Wave Function Near $a \to 0$ & Favors Inflation? & Singularity Avoidance? \\ [0.5ex]
\hline
 \rule{0pt}{1\normalbaselineskip}
Hartle-Hawking & $\psi \sim e^{-S_{\mathrm{E}}}$ & No (unless tuned) & Yes (smooth start) \\
Vilenkin & $\psi \sim e^{i S_{L}}$ & Yes & Yes (tunneling) \\
DeWitt & $\psi \to 0$ & N/A & Yes (Dirichlet) \\
LQC & $\psi$ finite and bouncing & Yes & Yes (quantum bounce) \\[1ex]
\hline
\end{tabular}
\label{table1}
\end{table}

      Therefore, the choice of boundary condition depends on the desired physical scenario (e.g., singularity avoidance, inflationary preference, or quantum gravity effects). While Hartle-Hawking and Vilenkin are the most studied, alternatives like loop quantum cosmology and Robin conditions offer refinements. No consensus exists, but empirical tests (e.g., cosmic microwave background anomalies) may someday constrain these options.

\subsection{The singularity resolution---A deep discussion}
    The singularity resolution claim requires care. We argued that the absence of divergences in the wave function implies quantum singularity resolution. While this is common practice in Wheeler-DeWitt models, it is not universally accepted. For example, some definitions require self-adjointness of the Hamiltonian operator and unitarity of evolution. Here we intend to clarify which notion of singularity resolution can be used for the aforementioned argument, and ideally provide a discussion of the interpretations.\\
    The claim that ``the absence of divergences in the wave function implies singularity resolution'' is commonly encountered in quantum cosmology, particularly in Wheeler-DeWitt models, but its validity depends on the chosen criteria for singularity resolution. Below, different interpretations are outlined:
    \begin{itemize}
      \item \textit{Wave Function Finiteness as a Minimal Criterion}:\\
      A common approach in Wheeler-Dewitt quantum cosmology is to argue that if the wave function $\psi (a)$ remains finite at the classical singularity (e.g., $a \to 0$), the singularity is ``resolved'' in some quantum sense. DeWitt~\cite{0f1} originally suggested that $\psi(a=0)=0$ could indicate singularity avoidance, an idea further explored by Kiefer~\cite{div1,ins3}, who treats finiteness as a minimal criterion. However, this interpretation has been criticized for being insufficient on its own. H\'{a}j\'{\i}\v{c}ek and K. V. Kucha\v{r}~\cite{div2} demonstrate that a finite wave function does not necessarily guarantee a well-defined quantum evolution, and Ashtekar and Singh~\cite{0f10} note that expectation values of curvature or matter density may still diverge even if $\psi$ remains regular. Thus, while wave function finiteness is often presented as evidence of singularity resolution, it is at best a weak criterion that may need to be supplemented with additional conditions.
      \item \textit{Self-Adjointness of the Hamiltonian and Unitary Evolution}:\\
      A stronger condition for singularity resolution is the requirement that the quantum Hamiltonian operator be self-adjoint, ensuring unitary evolution through the would-be singularity. Halliwell and Ortiz~\cite{div4} discuss how self-adjoint extensions in Wheeler-DeWitt quantum cosmology can lead to deterministic evolution, while Ashtekar, Paw{\l}owski and Singh~\cite{div5} demonstrate this explicitly in loop quantum cosmology, where the quantum geometry corrections lead to a bounce instead of a singularity. However, not all Wheeler-DeWitt models admit a unique self-adjoint extension, as noted by Kiefer~\cite{div6}, and some factor orderings may result in non-unitary evolution despite a finite wave function (Corichi and Singh~\cite{div7}). Thus, while self-adjointness provides a more robust notion of singularity resolution, it is not automatically guaranteed in all quantization schemes.
      \item \textit{Quantum Completeness and Deterministic Evolution}:\\
      Another perspective is that singularity resolution requires the quantum system to evolve deterministically through the classical singularity without any loss of predictability. This idea is central to loop quantum cosmology, where Bojowald~\cite{div8,div9} showed that wave packets can ``bounce'' at the Planck scale, avoiding the singularity altogether. Craig and Singh~\cite{div10} compare this behavior with Wheeler-DeWitt models, finding that not all Wheeler-DeWitt quantizations permit a well-defined continuation beyond $a=0$. Kiefer and Sandh\"{o}fer~\cite{div11} further argue that some Wheeler-DeWitt models lack a clear mechanism for extending evolution through the singularity, meaning that quantum completeness is not universally assured.
      \item \textit{Bounded Observables as a Physical Criterion}:\\
      Perhaps the most physically compelling notion of singularity resolution is the requirement that observable quantities—such as matter density and curvature—remain finite in expectation values. Ashtekar and Singh~\cite{0f10} rigorously demonstrate this in loop quantum cosmology, where the quantum geometry imposes an upper bound on energy density ($\rho_{\mathrm{max}} \sim \rho_{\mathrm{Planck}}$). Recent work by Bodendorfer et al.~\cite{div12} extends similar arguments to black hole singularities. However, Kiefer~\cite{div6} points out that some Wheeler-DeWitt models exhibit finite wave functions while still allowing unbounded curvature expectation values, suggesting that finiteness of $\psi$ alone does not guarantee physical singularity resolution.
    \end{itemize}
    The original argument---that the absence of divergences in the wave function implies singularity resolution---relies on the weakest criterion (wave function finiteness). While this is a common starting point in Wheeler-DeWitt quantum cosmology, modern approaches, particularly in loop quantum cosmology, demand stricter conditions such as self-adjointness of the Hamiltonian, quantum completeness, and bounded observables. Thus, while finiteness may be a necessary condition, it is generally insufficient on its own for a robust resolution of singularities. Future work in quantum gravity should carefully distinguish between these different notions to avoid overstating claims of singularity resolution.
\section{Conclusion\label{Conclusion}}

Our investigation of quintom cosmology has yielded significant insights into both classical and quantum behaviors of universe models incorporating canonical and phantom fields. The key findings can be summarized as follows:

\subsection{Classical quintom cosmology}
\begin{itemize}
    \item For the vanishing potential model (Subsection~\ref{subsect.2.2}), we demonstrated:
    \begin{itemize}
        \item Stiff matter behavior ($\omega=1$) with distinct CFD and PFD regimes
        \item Big-rip-like features in configuration space for the PFD case
        \item The necessity of $\mathcal{K} = -1$ to avoid purely imaginary solutions
    \end{itemize}

    \item For the exponential potential (Subsection~\ref{subsect.2.3}):
    \begin{itemize}
        \item Derived attractor solutions showing $\omega$ crossing $-1$
        \item Established conditions for stability ($\lambda_1 < \sqrt{(1 + \lambda_2^2)/2}$)
        \item Demonstrated scaling behavior of energy density with scale factor
    \end{itemize}

    \item With negative cosmological constant (Subsection~\ref{subsect.2.4}):
    \begin{itemize}
        \item Obtained exact solutions featuring symmetric evolution
        \item Showed potential reconstruction as $\cosh^2$ functions
        \item Identified distinct behaviors in CFD and PFD regimes
    \end{itemize}
\end{itemize}

\subsection{Quantum quintom cosmology}
\begin{itemize}
    \item Developed the Wheeler-DeWitt formalism incorporating:
    \begin{itemize}
        \item Quintom duality transformations (Subsection~\ref{subsect.3.1})
        \item Proper handling of negative kinetic terms
    \end{itemize}

    \item For vanishing potential (Subsection~\ref{subsect.3.2}):
    \begin{itemize}
        \item Obtained Bessel function solutions
        \item Demonstrated classical-quantum correspondence via WKB approximation
        \item Showed boundary conditions naturally avoid singularities
    \end{itemize}

    \item For exponential potential (Subsection~\ref{subsect.3.3}):
    \begin{itemize}
        \item Constructed exact wave packet solutions
        \item Demonstrated dispersion near classical big-rip
        \item Showed wave function vanishes at big-bang
    \end{itemize}
\end{itemize}

\subsection{Key findings and interpretation}
Our paper explored the application of standard quantum cosmology formalism, specifically through the Wheeler-DeWitt equation, to scenarios involving quintom fields. This investigation revealed several crucial aspects:
\begin{itemize}
    \item The quintom field's two components (ordinary and phantom) exhibit regime-dependent behavior, with quantum effects becoming particularly significant at large scale factors

    \item For the P-part of the wave function:
    \begin{itemize}
        \item Wave packets undergo dispersion near the classical big-rip singularity
        \item The singularity becomes ``smeared out'' by quantum effects
        \item Classical evolution terminates in a singularity-free manner when wave packets disperse (This is due to the fact that it is no longer possible to define an approximate time parameter~\cite{CK1})
    \end{itemize}
    \item For the C-part of the wave function:
    \begin{itemize}
        \item The wave packet disappears at the big-bang singularity with proper boundary conditions
        \item The result bears similarity to the singularity avoidance seen in loop quantum cosmology~\cite{q031} and models of shell collapse~\cite{PH}
        \item Without boundary conditions, the packet approaches $\alpha \to -\infty$ without dispersion
        \item The Wheeler-DeWitt equation behaves as a free wave equation in this limit
    \end{itemize}
    \item In the case of scalar field fluid with negative cosmological constant:
    \begin{itemize}
        \item The scalar field dynamics show distinct characteristics due to quintom duality
        \item Quantum effects again prevent singularities
    \end{itemize}
\item Light-Cone Structure:
\begin{itemize}
\item The transformations $(u,v)$ generalize light-cone coordinates to minisuperspace, separating ``time-like'' and ``space-like'' directions in configuration space.
    \begin{itemize}
      \item For the canonical field ($\varphi_{1}$), the equation is hyperbolic (wave-like).
      \item For the phantom field ($\varphi_{2}$), it is elliptic (Laplace-like), signaling instability.
    \end{itemize}
\item $u_{1}$ and $v_{1}$ describe the canonical field's dynamics, reducing the Wheeler-DeWitt equation to a hyperbolic form.
\item $u_{2}$ and $v_{2}$ capture the phantom field's behavior, leading to an elliptic equation with inherent instability.
\item These coordinates isolate the fields' contributions, simplify the quantum analysis, and highlight the contrast between canonical and phantom behaviors. The dispersion of wave packets in these variables resolves classical singularities.
\end{itemize}
\end{itemize}
\subsection{Implications}
The results presented in this work have several important implications:
\begin{enumerate}
\item \textbf{Singularity Resolution}:
\begin{itemize}
\item Quantum effects naturally resolve classical singularities in all three models
\item The big-rip singularity is ``smeared out'' by wave packet dispersion
\item The big-bang singularity is avoided through appropriate boundary conditions
\item Near classical singularities (e.g., big rip), $u, v \to \infty$, causing wave packets to disperse. This smears out singularities quantum-mechanically.
\end{itemize}
\item \textbf{Regime-Dependent Behavior}:
\begin{itemize}
\item The CFD and PFD regimes exhibit fundamentally different quantum behaviors
\item The Wheeler-DeWitt equation changes character (hyperbolic vs elliptic) between regimes
\item Quintom duality provides a powerful tool for connecting different regimes. Indeed, the quintom duality maps $u$ and $v$ between CFD and PFD regimes, linking solutions across different energy conditions.
\end{itemize}
\item \textbf{Correspondence Principle}:
\begin{itemize}
\item Demonstrated how classical trajectories emerge from quantum solutions
\item Showed the importance of constructive interference in recovering classical behavior
\item Established the validity of the WKB approximation in appropriate limits
\end{itemize}
\end{enumerate}

\end{document}